\documentclass[showpacs,pra]{revtex4}
\usepackage{graphicx}
\usepackage{bm}
\begin{document}

\title{Faddeev equations in one-dimensional problems with resonant
interactions}

\author{V. A. Yurovsky}

\affiliation{School of Chemistry, Tel Aviv University, 69978 Tel Aviv,
Israel}

\date{\today}
\begin{abstract}

A problem of three one-dimensional bosons with resonant multichannel
interactions is considered. The problem is reduced to a single-channel
Faddeev-Lovelace equation by elimination of the closed and output
 channels.
The equation is regularized using algebraic properties of their
singularities.

\end{abstract}
\maketitle

\section*{Introduction}

Quasi-one-dimensional (quasi-1D) quantum gases can be formed in
 prolongated potentials (atom
waveguides) whenever the atom energies are much less then the
 transverse excitation energies and only the
ground transverse state is substantially populated. Such gases have
 been recently realized in 2D optical
lattices
\cite{Greiner2001,Moritz2005,Kinoshita2004,Kinoshita2005,Kinoshita2006,Tolra2004,Fertig2005,Yukalov2009},
elongated atomic traps
\cite{Gorlitz2001,Leanhardt2002,Strecker2002,Khaykovich2002,Richard2003,Hugbart2007}, and atomic chips
\cite{Folman2002,Esteve2006,Hofferberth2007,Jo2007,Amerongen2008}.

Properties of quasi-1D gases are drastically different from the
3D ones (see review \cite{Yurovsky2008b} and the references therein)
Under certain conditions quasi-1D gases can be described by the
integrable Lieb-Liniger-McGuire (LLMG) model
\cite{Lieb1963,McGuire1964,Berezin1964} of 1D indistinguishable Bose
atoms with zero-range energy-independent interactions. This model has
an exact Bethe-ansatz solution and can describe di- and multi-atomic
molecules in the case of attractive interactions. However, reflection
and dissociation in atom-molecule collisions and three-atomic
association are forbidden within the LLMG model. Its high internal
symmetry can be lifted in more realistic systems, leading to
observable effects of integrability, such as atom-diatom reflection
for non-identical atoms \cite{Dodd1972}. Richer physical phenomena can
be provided by resonant interactions.

Resonances in interactions of quasi-1D atoms can arise both from
 virtual
excitation of transverse modes (confinement-induced resonance, or CIR,
\cite{Bergeman2003,Moore2004}) and from atomic internal structure
(Feshbach
resonance, see \cite{Timmermans1999}). Combined effects of Feshbach
 resonances
and CIR on two-atomic systems were analyzed in
 \cite{Yurovsky2005,Yurovsky2006b},
where parameters of 1D scattering were related to parameters of 3D
 scattering and
harmonic waveguide. Effects of an anharmonic confinement, finite-range
 interactions,
and other additional effects were analyzed in
\cite{Kim2005,Peano2005,Peano2005a,Kim2006,Naidon2007,Melezhik2007,Saeidian2008}.

Resonant interactions can lift integrability already in three-body
systems. CIR  can lead to atom-diatom reflection \cite{Mora2005} (a
 similar
process of atom-soliton reflection was analyzed in \cite{Sinha2006})
 and to
thermalization in three-atom collisions \cite{Mazets2008}. Narrow
 Feshbach
resonances can provide richer physical phenomena, such as association
 in
three-atom collisions, dissociation in atom-diatom collisions
\cite{Yurovsky2006} and stabilization of Feshbach molecules
\cite{Yurovsky2008}.

An effective approach to three-body problems was developed on the base
of Faddeev equations (see books \cite{Schmid,Glockle} and the
 references
therein). Exact analytical \cite{Dodd1970,Majumdar1972} and numerical
\cite{Mehta2005} solutions of Faddeev equations for the LLMG model
 agree
with the Bethe-ansatz solution. Numerical solution of Faddeev-Lovelace
equations has been applied in \cite{Dodd1972} to asymmetric systems
 and in
\cite{Melde2002} to 1D systems with generic separable interactions. An
equivalent method was used in \cite{Mora2005}. An alternative method
 of
numerical solution of the Schr\"odinger equation in hyperspherical
coordinates was applied to 1D problems in \cite{Mehta2007}. Three-body
bound states for resonant interactions were analyzed in
\cite{Kheruntsyan1998} using a variational approach.

The present work gives a comprehensive description of the Faddeev
approach to 1D problems, generalizing the approach of
\cite{Yurovsky2006,Yurovsky2008} to {\it multichannel} resonant
 interactions.
The second-quantized Hamiltonian for this case is introduced in Sec.\
\ref{Hamiltonian}. Two-body scattering and bound states are analyzed
 in
Sec.\ \ref{TwoBody}. Section \ref{ThreeBody} describes the three-body
multichannel problem. Asymptotic states and transition amplitudes for
 this
problem are introduced in Sec.\ \ref{MultFadd}. Section
 \ref{SingChann}
relates the multichannel transition amplitudes to solutions of the
Faddeev-Lovelace equation for an effective single-channel problem.
Probabilities of bound-bound and bound-free transitions are presented
 in
Sec.\ \ref{TranProb}. Section \ref{PrivCase} describes special cases
 of
the two-channel system and deactivation problem. The solution method
 of
Faddeev-Lovelace equations is described in Sec.\ \ref{SolFaddLov}. The
present regularization procedure is applicable to generic 1D problems
 with
separable interactions. The singularities related to the poles of
 two-body
$T$ matrix, which are similar to the 3D case, are treated here using a
generalization of the approach \cite{Dodd1972}. The present approach
 also
regularizes the singularities of the free-atom Green function, which
 are
different from the 3D case. They appear in the processes of
 dissociation
and association and were not considered in \cite{Dodd1972}.
 Derivation of
certain properties of resonant two- and three-body models is included
 in
the Appendices.

\section{One-dimensional bosons with multichannel zero-range
interactions}\label{1Dbosons}

\subsection{Hamiltonian}\label{Hamiltonian}

Consider a gas of 1D Bose atoms described by the annihilation
operators $\hat{\Psi }_{a}\left( z\right) $. The model includes
 several two-body channels described by
the annihilation operators $\hat{\Psi }_{m}\left( z\right) $ of
 molecules with the energies $D_{m}$. The
Hamiltonian of the system has the form (using units with Plank's
 constant
$\hbar =1$),
\begin{eqnarray}
\hat{H}=\int dz\biggl\{\hat{\Psi }^{\dag }_{a}\left( z\right)
 \left\lbrack -{1\over 2m} {\partial { } ^{2}\over \partial z{ }
 ^{2}}+{U{ } _{a}\over 2}\hat{\Psi }^{\dag }_{a}\left( z\right)
 \hat{\Psi }_{a}\left( z\right) \right\rbrack \hat{\Psi }_{a}\left(
 z\right)  \nonumber
\\
+\sum\limits^{}_{m}\hat{\Psi }^{\dag }_{m}\left( z\right) \left(
 -{1\over 4m}{\partial { } ^{2}\over \partial z{ } ^{2}}+D_{m}\right)
 \hat{\Psi }_{m}\left( z\right) +\sum\limits^{}_{m}\left\lbrack
 \hat{V}_{am}\left( z\right) +\hat{V}^{\dag }_{am}\left( z\right)
 \right\rbrack +\sum\limits^{}_{m\neq m^\prime }\hat{V}_{mm^\prime
 }\biggr\} , \label{Hsq}
\end{eqnarray}
where $m$ is the atomic mass and $U_{a}$  is the non-resonant
 interatomic
interaction strength. At negative values of $U_{a}$  the open
 channel, formed by
two atoms in the initial state $\hat{\Psi }_{a}$, has a bound state
 too. The interactions
\begin{equation}
\hat{V}_{am}\left( z\right) =g_{m}\hat{\Psi }^{\dag }_{m}\left(
 z\right) \hat{\Psi }_{a}\left( z\right) \hat{\Psi }_{a}\left(
 z\right)
\end{equation}
couple the open channel to other (molecular) channels and
\begin{equation}
\hat{V}_{mm^\prime }\left( z\right) =d_{mm^\prime }\hat{\Psi }^{\dag
 }_{m}\left( z\right) \hat{\Psi }_{m^\prime }\left( z\right)  ,\qquad
 d_{m^\prime m}=d^{*}_{mm^\prime } ,
\end{equation}
couple the molecular channels [the matrix $d_{m^\prime m}$  will be
 completed by
diagonal elements, see Eq.\ (\ref{diagdmmp}) below]. The system with a
single molecular state was analyzed in
 \cite{Yurovsky2005,Yurovsky2006b},
where the one-dimensional parameters were related to the atomic
 collision
and waveguide parameters. Additional molecular states with large
 negative
$D_{m}$  were introduced in \cite{Yurovsky2008}.

\subsection{Two-body system} \label{TwoBody}

A state vector of the two-atom system can be represented as a
superposition of atomic and molecular states,
\begin{equation}
|\Psi _{2}\rangle =\int dz_{c}e^{iPz_c}\biggl\lbrack {1\over
 \sqrt{2}}\int\limits^{\infty }_{-\infty }dz \varphi ^{\left( 0\right
) }\left( z\right) \hat{\Psi }^{\dag }_{a}\left( z_{c}-{z\over
 2}\right) \hat{\Psi }^{\dag }_{a}\left( z_{c}+{z\over 2}\right)
+\sum\limits^{}_{m} \varphi ^{\left( m\right) }\hat{\Psi }^{\dag
 }_{m}\left( z_{c}\right)  \biggr\rbrack |\text{vac}\rangle  ,
 \label{Psi2}
\end{equation}
where $z_{c}$  and $P$ are the center-of-mass position and momentum,
respectively, $z$ is the interatomic distance, and $|$vac$\rangle $
 is the physical
vacuum state. Substitution of the state vector (\ref{Psi2}) into the
Schr\"odinger equation with the Hamiltonian (\ref{Hsq}) leads to the
coupled equations for the channel functions $\varphi ^{\left( 0\right
) }\left( z\right) $ and $\varphi ^{\left( m\right) }$,
\begin{eqnarray}
E\varphi ^{\left( 0\right) }\left( z\right) =\left\lbrack -{1\over m}
 {d{ } ^{2}\over dz{ } ^{2}}+U_{a}\delta \left( z\right)
 \right\rbrack \varphi ^{\left( 0\right) }\left( z\right)
+\sqrt{2}\delta \left( z\right) \sum\limits^{}_{m}g^{*}_{m}\varphi
 ^{\left( m\right) } \label{cchan20}
\\
E\varphi ^{\left( m\right) }=D_{m}\varphi ^{\left( m\right) }
+\sqrt{2}g_{m}\varphi ^{\left( 0\right) }\left( 0\right)
+\sum\limits^{}_{m^\prime \neq m}d_{mm^\prime }\varphi ^{\left(
 m^\prime \right) } \label{cchan2m}
\end{eqnarray}
Here $E$ is the energy in the center-of-mass system.

The molecular channel functions can be expressed as
\begin{equation}
\varphi ^{\left( m\right) }=\sqrt{2}\varphi ^{\left( 0\right) }\left(
 0\right) G_{m}\left( E\right)  ,
\end{equation}
where the vector $G_{m}$  is a solution of the system of linear
equations
\begin{equation}
\sum\limits^{}_{m^\prime }d_{mm^\prime }\left( E\right) G_{m^\prime
 }\left( E\right) =-g_{m} \label{Gmeq}
\end{equation}
and the diagonal elements of the matrix $d_{mm^\prime }\left( E\right
) $ are defined as
\begin{equation}
d_{mm}\left( E\right) =D_{m}-E-i0 . \label{diagdmmp}
\end{equation}
If $g_{m}=-\delta _{mm^\prime }$, then $G_{m}\left( E\right) $ has a
 physical sense of a Green
function, and the infinitesimal imaginary part of $d_{mm}\left(
 E\right) $ specifies
its retarding behavior.

Elimination of the molecular channels leads to a single equation
for the open-channel function,
\begin{equation}
E\varphi ^{\left( 0\right) }\left( z\right) =\left\lbrack -{1\over m}
 {d{ } ^{2}\over dz{ } ^{2}}+U_{\text{eff}}\left( E\right) \delta
 \left( z\right) \right\rbrack \varphi ^{\left( 0\right) }\left(
 z\right)  , \label{phi0Ueff}
\end{equation}
where the effective energy-dependent interaction strength
\begin{equation}
U_{\text{eff}}\left( E\right) =U_{a}+2\sum\limits^{}_{m}g^{
*}_{m}G_{m}\left( E\right)  . \label{Ueff}
\end{equation}
incorporates effects of all channels. In the case of a single
molecular channel it is reduced to the expression for
 $U_{\text{eff}}\left( E\right) $ in
\cite{Yurovsky2005}.

Obviously, the wavefunction of a two-body scattering state can be
expressed as
\begin{equation}
\varphi ^{\left( 0\right) }\left( z\right) =\left( 2\pi \right) ^{-1
/2}\left\lbrack e^{ikz}-i{m\over 2k}T_{1D}\left( k\right) e^{ik|z
|}\right\rbrack
\end{equation}
in terms of the two-body $T$ matrix
\begin{equation}
T_{1D}\left( k\right) =\left\lbrack U^{-1}_{\text{eff}}\left( k^{2}
/m\right) +{i\over 2}mk^{-1}\right\rbrack ^{-1} , \label{T1D}
\end{equation}
which depends on the relative momentum $k$ of two colliding atoms
($E=k^{2}/m$). $T_{1D}\left( k\right) $  has poles on the positive
 imaginary axis, $k=i\kappa _{n}$, where $\kappa _{n}$
are solutions of the equation
\begin{equation}
\kappa =-{m\over 2}U_{\text{eff}}\left( -{\kappa { } ^{2}\over
 m}\right)  . \label{kappan}
\end{equation}
The poles correspond to two-body bound states (diatoms) with the
energies $E_{n}=-\kappa ^{2}_{n}/m$. The diatom states are orthogonal
 (see App.\
\ref{AppOrt}) and the orthonormality conditions have the form
\begin{equation}
\int\limits^{\infty }_{-\infty }dz\varphi _{n^\prime }^{(0)*}\left(
 z\right) \varphi ^{\left( 0\right) }_{n}\left( z\right)
+\sum\limits^{}_{m}\varphi ^{\left( m\right) *}_{n^\prime }\varphi
 ^{\left( m\right) }_{n}=\delta _{nn^\prime } . \label{DiatomOrt}
\end{equation}
\medskip
Solutions of Eq.\ (\ref{Gmeq}) can be expressed in terms of
 $\det\left( d\right) $,
the determinant of the matrix $d_{mm^\prime }\left( \kappa ^{2}
/m\right) $, which is a polynomial of the
degree $2M_{\text{mol}}$  in $\kappa $, and the cofactors $\partial
 \det\left( d\right) /\partial d_{mm^\prime }$, which are polynomials
of the degree $2\left( M_{\text{mol}}-1\right) $ in $\kappa $ (where
 $M_{\text{mol}}$  is the number of the molecular
channels). Using Eq.\ (\ref{Ueff}), Eq.\ (\ref{kappan})  can be
 represented
then in the polynomial form,
\begin{equation}
\left( {2\over m}\kappa +U_{a}\right) \det\left( d\right)
 -\sum\limits^{}_{m,m^\prime }g_{m^\prime }{\partial \det\left(
 d\right) \over \partial d{ } _{mm^\prime }}g^{*}_{m}=0 ,
 \label{kappanpol}
\end{equation}
This equation has $2M_{\text{mol}}+1$ roots (only the real positive
 roots
correspond to the bound states). Thus the energy-dependent
 interaction can
lead to multiple bound states in spite of its zero range. The diatoms
 are
superpositions of the open and molecular channels. The open-channel
component can be expressed as
\begin{equation}
\varphi ^{\left( 0\right) }_{n}\left( z\right) =\varphi ^{\left(
 0\right) }_{n}\left( 0\right) \exp\left( -\kappa _{n}|z|\right)  ,
 \label{phin0}
\end{equation}
where $\varphi ^{\left( 0\right) }_{n}(0)$  is determined by the
 normalization conditions
(\ref{DiatomOrt}), and the open-channel contribution to the bound
 state is
\begin{equation}
W_{n}=\int\limits^{\infty }_{-\infty }dz |\varphi ^{\left( 0\right)
 }_{n}\left( z\right) |^{2}=\left( 1+2\sum\limits^{}_{m}|G_{m}\left(
 E_{n}\right) |^{2}\kappa _{n}\right) ^{-1} . \label{Wn}
\end{equation}
Although the present model approximates the molecular-channel states
to be infinitesimal in size, the diatoms have finite sizes ($\sim
 \kappa ^{-1}_{n}$).

\subsection{Three-body systems}\label{ThreeBody}

A system of three atoms is described by the state vector
\begin{equation}
|\Psi _{3}\rangle =\biggl\lbrack {1\over \sqrt{6}}\int d^{3}z\psi
 ^{\left( 0\right) }\left( z_{1},z_{2},z_{3}\right) \hat{\Psi }^{\dag
 }_{a}\left( z_{1}\right) \hat{\Psi }^{\dag }_{a}\left( z_{2}\right)
 \hat{\Psi }^{\dag }_{a}\left( z_{3}\right) +\sum\limits^{}_{m}\int
 dz dz_{m}\psi ^{\left( m\right) }\left( z,z_{m}\right) \hat{\Psi
 }^{\dag }_{a}\left( z\right) \hat{\Psi }^{\dag }_{m}\left(
 z_{m}\right) \biggr\rbrack |\text{vac}\rangle  , \label{Psi3}
\end{equation}
leading to the coupled equations for the wavefunctions of the
three-atom $\psi ^{\left( 0\right) }\left( z_{1},z_{2},z_{3}\right) $
 and atom-molecule $\psi ^{\left( m\right) }\left( z,z_{m}\right) $
 channels. Their
momentum representations, $\tilde{\psi }^{\left( 0\right) }\left(
 q,k\right) $ and $\tilde{\psi }^{\left( m\right) }\left( q\right) $,
 are defined by
\begin{eqnarray}
\psi ^{\left( 0\right) }\left( z_{1},z_{2},z_{3}\right) =\left( 2\pi
 \right) ^{-3/2}\int dq_{j}dk_{j}\tilde{\psi }^{\left( 0\right)
 }\left( q_{j},k_{j}\right) \exp\left( iq_{j}\left(
 z_{j}-{z_{j^\prime +}z{ } _{j^{\prime\prime}}\over 2}\right)
+ik_{j}\left( z_{j^\prime }-z_{j^{\prime\prime}}\right) +iP{z_{1}
+z_{2}+z{ } _{3}\over 3}\right)  \nonumber
\\
\label{psi3} ,
\\
\psi ^{\left( m\right) }\left( z,z_{m}\right) =\left( 2\pi \right)
 ^{-1}\int dq\tilde{\psi }^{\left( m\right) }\left( q_{j}\right)
 \exp\left( iq_{j}\left( z-z_{m}\right) +iP{z+2z{ } _{m}\over
 3}\right)  \nonumber
\end{eqnarray}
where $P$ is center-of-mass momentum and $q_{j},k_{j}$  are momenta
 in any
of the three sets of Jacobi coordinates ($j=1,2,3$). The momentum
 $q_{j}$  is
the relative momentum of the $j$ th atom and the center-of-mass of the
$j^\prime $ th and $j^{\prime\prime}$ th atoms, while $k_{j}$  is the
 relative momentum of the $j^\prime $ th
and $j^{\prime\prime}$ th atoms, where $j$, $j^\prime $, and
 $j^{\prime\prime}$  are cyclic permutations of 1, 2,
and 3. The momenta in different Jacobi coordinate sets are related as
\begin{equation}
q_{j^\prime ,j^{\prime\prime}}=-{1\over 2}q_{j}\pm k_{j} ,\qquad
 k_{j^\prime ,j^{\prime\prime}}=\mp {3\over 4}q_{j}+{1\over 2}k_{j} .
 \label{JacRel}
\end{equation}
These relations correspond to permutations of the atoms. In what
follows the center-of-mass system ($P=0$) is used.

The channel wavefunctions in the momentum representation satisfy
the following equations
\begin{eqnarray}
E\tilde{\psi }^{\left( 0\right) }\left( q_{j},k_{j}\right) =\left(
 {3\over 4m}q^{2}_{j}+{1\over m}k^{2}_{j}\right) \tilde{\psi }^{\left
( 0\right) }\left( q_{j},k_{j}\right) +{U{ } _{a}\over 2\pi
 }\sum\limits^{3}_{l=1} \int dk^\prime _{l}\tilde{\psi }^{\left(
 0\right) }\left( q_{l},k_{l}^\prime \right) +\left( 3\pi \right)
 ^{-1/2}\sum\limits^{}_{m}g^{*}_{m}\sum\limits^{3}_{l=1}\tilde{\psi
 }^{\left( m\right) }\left( q_{l}\right)  \nonumber
\\
\label{cchan3m} ,
\\
E\tilde{\psi }^{\left( m\right) }\left( q_{j}\right) =\left( {3\over
 4m}q^{2}_{j}+D_{m}\right) \tilde{\psi }^{\left( m\right) }\left(
 q_{j}\right) +\left( {3\over \pi }\right) ^{1/2}g_{m}\int
 dk_{j}\tilde{\psi }^{\left( 0\right) }\left( q_{j},k_{j}\right)
+\sum\limits^{}_{m^\prime \neq m}d_{mm^\prime }\tilde{\psi }^{\left(
 m^\prime \right) }\left( q_{j}\right)  \nonumber
\end{eqnarray}
where $E$ is the energy in the center-of-mass system. These equations
are invariant over transformations (\ref{JacRel}) and can describe
indistinguishable Bose atoms.

Equations (\ref{cchan3m}) can be represented in a compact vector form.
Consider for the moment distinguishable atoms and introduce the vector
wavefunction $\tilde{\bm{\psi }}=\{\tilde{\psi }^{\left( \alpha
 \right) }\}$, where $\alpha $ has values 0, $m1$, $m2$, and $m3$
for all $m$. The component $\tilde{\psi }^{\left( mj\right)
 }=\tilde{\psi }^{\left( m\right) }\left( q_{j}\right) $ describes
 the atom-molecule channel
with free $j$ th atom. Equations (\ref{cchan3m}) can be then written
 out in
the form
\begin{equation}
E\tilde{\bm{\psi }}=\left( \hat{\bm{H}}_{0}
+\sum\limits^{3}_{l=1}\hat{\bm{U}}_{l}\right) \tilde{\bm{\psi }} ,
 \label{vect3b}
\end{equation}
where elements of the matrices $\hat{\bm{H}}_{0}$  and
 $\hat{\bm{U}}_{l}$  are
defined, respectively, as
\begin{eqnarray}
\hat{H}^{\alpha ^\prime \alpha }_{0}=\left\lbrack \left( {3\over
 4m}q^{2}_{j}+{1\over m}k^{2}_{j}\right) \delta _{\alpha 0}
+\sum\limits^{}_{m,j}\left( {3\over 4m}q^{2}_{j}+D_{m}\right) \delta
 _{\alpha ,mj}\right\rbrack \delta _{\alpha \alpha ^\prime }
\\
\hat{U}^{\alpha ^\prime \alpha }_{l}\left\lbrack \tilde{\psi }^{\left
( \alpha \right) }\right\rbrack =\left\lbrack {U{ } _{a}\over 2\pi
 }\delta _{\alpha ^\prime 0}+\left( {3\over \pi }\right) ^{1
/2}\sum\limits^{}_{m,j}g_{m}\delta _{\alpha ^\prime ,mj}\delta
 _{lj}\right\rbrack \delta _{\alpha 0}\int dk^\prime _{l}\tilde{\psi
 }^{\left( 0\right) }\left( q_{l},k^\prime _{l}\right)  \nonumber
\\
+\sum\limits^{}_{m,j}\left\lbrack \left( 3\pi \right) ^{-1/2}g^{
*}_{m}\delta _{\alpha ^\prime 0}+\sum\limits^{}_{m^\prime ,j^\prime
 }d_{m^\prime m}\delta _{\alpha ^\prime ,m^\prime j^\prime }\delta
 _{lj^\prime }\right\rbrack \delta _{\alpha ,mj}\delta
 _{lj}\tilde{\psi }^{\left( mj\right) }\left( q_{j}\right)
 \label{Umchan}
\end{eqnarray}
The definition of $\hat{H}^{\alpha \alpha ^\prime }_{0}$  is
 unambiguous since the kinetic energy
${3\over 4m}q^{2}_{j}+{1\over m}k^{2}_{j}$  keeps the same value in
 the three Jacobi coordinate sets.

The atom-molecule channel wavefunctions can be expressed in terms
of the three-atom one,
\begin{equation}
\tilde{\psi }^{\left( m\right) }\left( q_{j}\right) =\left( {3\over
 \pi }\right) ^{1/2}G_{m}\left( E-{3\over 4m}q^{2}_{j}\right) \int
 dk_{j}\tilde{\psi }^{\left( 0\right) }\left( q_{j},k_{j}\right)  ,
 \label{psimom}
\end{equation}
where $G_{m}$  is defined by Eq.\ (\ref{Gmeq}). Elimination of the
atom-molecule channels leads to a single equation for the three-atom
wavefunction,
\begin{equation}
E\tilde{\psi }^{\left( 0\right) }\left( q_{j},k_{j}\right) =\left(
 {3\over 4m}q^{2}_{j}+{1\over m}k^{2}_{j}\right) \tilde{\psi }^{\left
( 0\right) }\left( q_{j},k_{j}\right)
+\sum\limits^{3}_{l=1}\hat{U}^{\text{eff}}_{l}\left\lbrack
 \tilde{\psi }^{\left( 0\right) }\right\rbrack  , \label{EqpsiUeff}
\end{equation}
where the interaction operators are defined as
\begin{equation}
\hat{U}^{\text{eff}}_{l}\left\lbrack \tilde{\psi }^{\left( 0\right)
 }\right\rbrack ={1\over 2\pi }U_{\text{eff}}\left( E-{3\over
 4m}q^{2}_{l}\right) \int dk^\prime _{l}\tilde{\psi }^{\left( 0\right
) }\left( q_{l},k^\prime _{l}\right)
\end{equation}
with $U_{\text{eff}}$  given by Eq.\ (\ref{Ueff}). Elimination of
 channels and
energy-dependent potentials in 3D three-body problems were considered
 in
\cite{Abdurakhmanov1985,Abdurakhmanov1987,Vinitskii1992}.

However, in three-body problems effects of the eliminated channels
extend beyond the interaction strength. The additional effects are
analyzed below.

\subsection{Multichannel Faddeev approach}\label{MultFadd}

Equation (\ref{vect3b}) describes a vector three-body problem. Similar
scalar problems were extensively studied using Faddeev equations (see
 books
\cite{Schmid,Glockle}). In line with this approach let us introduce
 vector
functions $\tilde{\bm{\chi }}_{0{\bf Q}}$  and $\tilde{\bm{\chi
 }}_{lnp}$, describing the
asymptotic channels. The wavefunction of three free atoms with the
 momenta
$Q_{1}$, $Q_{2}$, and $Q_{3}$, such that $Q_{1}+Q_{2}+Q_{3}=P=0$, has
 the form
\begin{equation}
\tilde{\chi }^{\left( \alpha \right) }_{0{\bf Q}}\left(
 q_{j},k_{j}\right) =\delta _{\alpha 0}\delta \left(
 q_{j}-Q_{j}\right) \delta \left( k_{j}-{Q_{j^\prime }-Q{ }
 _{j^{\prime\prime}}\over 2}\right)  \label{chiv0Q}
\end{equation}
in each of the Jacobi coordinate sets. The asymptotic function
containing  the free $l$ th atom and a diatom in the state $n$ is
\begin{equation}
\tilde{\chi }^{\left( 0\right) }_{lnp}\left( q_{l},k_{l}\right)
 =\delta \left( q_{l}-p\right) \tilde{\varphi }^{\left( 0\right)
 }_{n}\left( k_{l}\right) ,\qquad \tilde{\chi }^{\left( mj\right)
 }_{lnp}\left( q_{l}\right) =\delta _{lj}\delta \left( q_{l}-p\right)
 \varphi ^{\left( m\right) }_{n} , \label{chimlnp}
\end{equation}
where $p$ is the relative momentum of the atom and diatom and
\begin{equation}
\tilde{\varphi }^{\left( 0\right) }_{n}\left( k\right) =\left(
 {2\kappa ^{3}_{n}W{ } _{n}\over \pi }\right) ^{1/2}{1\over k^{2}
+\kappa { } ^{2}_{n}} \label{phi0mom}
\end{equation}
is the momentum representation of the two-body wavefunction
(\ref{phin0}). The asymptotic functions satisfy the following
 equations
\begin{eqnarray}
E\tilde{\bm{\chi }}_{0{\bf Q}}=\hat{\bm{H}}_{0}\tilde{\bm{\chi
 }}_{0{\bf Q}} \nonumber
\\
{} ,
\\
E\tilde{\bm{\chi }}_{lnp}=\left( \hat{\bm{H}}_{0}
+\hat{\bm{U}}_{l}\right) \tilde{\bm{\chi }}_{lnp} \nonumber
\end{eqnarray}
where $E={\bf Q}^{2}/\left( 2m\right) $ in the first equation and
 $E=3p^{2}/\left( 4m\right) -\kappa ^{2}_{n}/m$   in the
second one.

The elements of the scattering matrix can be expressed as
\begin{eqnarray}
S_{l^\prime n^\prime p^\prime ,lnp}=\delta \left( p^\prime -p\right)
 \delta _{nn^\prime }\delta _{ll^\prime }-2\pi i\delta \left(
 3{p^{\prime 2}-p^2\over 4m}-{\kappa ^{2}_{n^\prime }-\kappa { }
 ^{2}_{n}\over m}\right) X_{l^\prime n^\prime p^\prime ,lnp} \nonumber
\\
{}
\\
S_{0{\bf Q},lnp}=-2\pi i\delta \left( {{\bf Q}{ } ^{2}\over 2m}-3{p{
 } ^{2}\over 4m}+{\kappa { } ^{2}_{n}\over m}\right) X_{0{\bf Q},lnp}
 \nonumber
\end{eqnarray}
in terms of the transition amplitudes
\begin{eqnarray}
X_{l^\prime n^\prime p^\prime ,lnp}=\langle \tilde{\bm{\chi
 }}_{l^\prime n^\prime p^\prime }
|\sum\limits^{}_{l^{\prime\prime}\neq l^\prime
 }\hat{\bm{U}}_{l^{\prime\prime}}|\tilde{\bm{\psi }}_{lnp}\rangle
 \label{Xmchan}
\\
X_{0{\bf Q},lnp}=\langle \tilde{\bm{\chi }}_{0{\bf Q}}
|\sum\limits^{3}_{l^{\prime\prime}=1}\hat{\bm{U}}_{l^{\prime\prime}}
|\tilde{\bm{\psi }}_{lnp}\rangle  \label{X0Qmchan}
\end{eqnarray}
Here the scattering-state solutions $\tilde{\bm{\psi }}_{lnp}$  of
 Eq.\
(\ref{vect3b}) have $\tilde{\bm{\chi }}_{lnp}$  as the incident wave.

Given energy $E$ and internal state $n$, the momentum $p$ can have
 values
$\pm p_{n}$, where
\begin{equation}
p_{n}=2\sqrt{\left( mE+\kappa ^{2}_{n}\right) /3} . \label{pn}
\end{equation}
The bound-bound $S$ matrix elements can be then represented as
\begin{equation}
S_{l^\prime n^\prime p^\prime ,ln\pm p_n}=\delta \left( p^\prime \mp
 p_{n}\right) \delta _{nn^\prime }\delta _{ll^\prime }-{4\pi im\over
 3p{ } _{n^\prime }}\left\lbrack \delta \left( p^\prime -p_{n^\prime
 }\right) +\delta \left( p^\prime +p_{n^\prime }\right) \right\rbrack
 X_{l^\prime n^\prime p^\prime ,ln\pm p_n} ,
\end{equation}
describing transmission and reflection in atom-diatom collisions with
possible rearrangement of atoms and change of the diatom internal
 state.

Indistinguishable Bose atoms are described by the wavefunctions
$\psi ^{\left( 0\right) }\left( z_{1},z_{2},z_{3}\right) $ which are
 independent over permutations of each pair of the
atomic coordinates $z_{1}$, $z_{2}$, and $z_{3}$. Then the
 momentum-representation
wavefunctions $\tilde{\psi }^{\left( 0\right) }\left(
 q_{j},k_{j}\right) $, $\tilde{\psi }^{\left( m\right) }\left(
 q_{j}\right) $ are invariant over transformations
(\ref{JacRel}), and the components $\tilde{\psi }^{\left( mj\right)
 }$  are independent of $j$. A symmetric
scattering-state solution $\tilde{\bm{\psi }}_{np}=3^{-1
/2}\sum\limits^{}_{l}\tilde{\bm{\psi }}_{lnp}$  is
obtained with a symmetrized incident wave $3^{-1
/2}\sum\limits^{}_{l}\tilde{\bm{\chi }}_{lnp}$  since
Eqs.\ (\ref{cchan3m}) and (\ref{vect3b}) are invariant over
 transformations
(\ref{JacRel}). Then a symmetric atom-diatom collision wavefunction
 in the
coordinate representation [see Eq.\ (\ref{psi3})] has the asymptotic
\begin{eqnarray}
\bm{\psi }_{np_n}\sim 3^{-1/2}\sum\limits^{3}_{l=1}\biggl\lbrack
 \bm{\chi }_{lnp_n}-{4\pi im\over 3}\sum\limits^{}_{n^\prime
 ,j}{1\over p{ } _{n^\prime }}\left( X_{jn^\prime p_{n^\prime
 },lnp_n}\bm{\chi }_{jn^\prime p_{n^\prime }}+X_{jn^\prime
 -p_{n^\prime },lnp_n}\bm{\chi }_{jn^\prime -p_{n^\prime }}\right)
 \nonumber
\\
-2\pi i\int d^{3}Q\delta \left( Q_{1}+Q_{2}+Q_{3}\right) \delta \left
( {{\bf Q}{ } ^{2}\over 2m}-E\right) X_{0{\bf Q},lnp_n}\bm{\chi
 }_{0{\bf Q}}\biggr\rbrack  , \label{psias}
\end{eqnarray}
where the three terms in the square brackets correspond to the
 incident
wave, transmission-reflection, and dissociation of the diatom,
 respectively,
and $\bm{\chi }$ are the coordinate representations of the functions
$\tilde{\bm{\chi }}$.

\subsection{Effective single-channel problem}\label{SingChann}

Taking into account Eqs.\ (\ref{Umchan}) and (\ref{chimlnp}), and
expressing $\tilde{\psi }^{\left( mj\right) }_{lnp}$  in terms of
 $\tilde{\psi }^{\left( 0\right) }_{lnp}$   using Eq.\ (\ref{psimom}
), one can
represent the transition amplitudes in the form of
\begin{eqnarray}
X_{l^\prime n^\prime \pm p_{n^\prime },lnp_n}=\langle \tilde{\chi
 }^{\left( 0\right) }_{l^\prime n^\prime \pm p_{n^\prime }}
|\sum\limits^{}_{l^{\prime\prime}\neq l^\prime
 }\hat{U}^{\text{eff}}_{l^{\prime\prime}}|\tilde{\psi }^{\left(
 0\right) }_{lnp_n}\rangle  \label{XUeff0}
\\
X_{0{\bf Q},lnp_n}=\langle \tilde{\chi }^{\left( 0\right) }_{0{\bf
 Q}}| \sum\limits^{3}_{l^{\prime\prime}=1}
 \hat{U}^{\text{eff}}_{l^{\prime\prime}}| \tilde{\psi }^{\left(
 0\right) }_{lnp_n}\rangle  ,
\end{eqnarray}
involving only the three-atom components of the wavefunctions. These
components are proportional to solutions of the effective
 single-channel
three-body problem with interaction $\hat{U}^{\text{eff}}_{l}$.
 Indeed, consider the solution of
the scalar equation
\begin{equation}
E\tilde{\chi }_{lnp}\left( q_{l},k_{l}\right)
 =\hat{H}^{00}_{0}\tilde{\chi }_{lnp}\left( q_{l},k_{l}\right)
+\hat{U}^{\text{eff}}_{l}\left\lbrack \tilde{\chi
 }_{lnp}\right\rbrack  ,
\end{equation}
of the form
\begin{equation}
\tilde{\chi }_{lnp}\left( q_{l},k_{l}\right) =\left( {2\kappa { }
 ^{3}_{n}\over \pi }\right) ^{1/2}{\delta \left( q_{l}-p\right) \over
 {3\over 4}q^{2}_{l}+k^{2}_{l}-mE} , \label{chilnp}
\end{equation}
where only the normalization factor depends on $n$ for fixed $p$. The
solutions are normalized as $\langle \tilde{\chi }_{lnp^\prime }
|\tilde{\chi }_{lnp}\rangle =A\delta \left( p^\prime -p\right) $,
 where $A=1$ for $p=p_{n}$.
Taking into account Eqs.\ (\ref{chimlnp}), (\ref{phi0mom}), and
(\ref{pn})), the wavefunctions in Eq.\ (\ref{XUeff0}) can be related
 to the
single-channel solutions for an arbitrary diatom state $n=0$, chosen
 as the
reference state,
\begin{eqnarray}
\tilde{\chi }^{\left( 0\right) }_{lnp_n}\left( q_{l},k_{l}\right)
 =\left( {W_{n}\kappa { } ^{3}_{n}\over \kappa { } ^{3}_{0}}\right)
 ^{1/2}\tilde{\chi }_{l0p_n}\left( q_{l},k_{l}\right) \label{chi0chi}
 \nonumber
\\
{}
\\
\tilde{\psi }^{\left( 0\right) }_{lnp_n}\left( q_{j},k_{j}\right)
 =\left( {W_{n}\kappa { } ^{3}_{n}\over \kappa { } ^{3}_{0}}\right)
 ^{1/2}\tilde{\psi }_{l0p_n}\left( q_{j},k_{j}\right)  .
 \label{psi0psi} \nonumber
\end{eqnarray}
Here $\tilde{\psi }_{l0p_n}$  is the scattering-state solution of Eq.\
(\ref{EqpsiUeff}) with the incident wave $\tilde{\chi }_{l0p_n}\left(
 q_{l},k_{l}\right) $ which is off-shell
[$3p^{2}_{n}/\left( 4m\right) +\kappa ^{2}_{0}/m\neq E$ for $n\neq
 0$]. Therefore, all transition amplitudes in Eq.\
(\ref{psias}) can be related to off-shell amplitudes for the effective
single-channel interaction $\hat{U}^{\text{eff}}_{l}$  and a single
 diatom state $n=0$ as,
\begin{eqnarray}
X_{l^\prime n^\prime \pm p_{n^\prime },lnp_n}=\sqrt{W_{n^\prime }W{ }
 _{n}}{\left( \kappa _{n^\prime }\kappa _{n}\right) { } ^{3/2}\over
 \kappa { } ^{3}_{0}}\langle \tilde{\chi }_{l^\prime 0\pm p_{n^\prime
 }}|\sum\limits^{}_{l^{\prime\prime}\neq l^\prime
 }\hat{U}^{\text{eff}}_{l^{\prime\prime}}|\tilde{\psi
 }_{l0p_n}\rangle  \nonumber
\\
\label{XUeff}
\\
X_{0{\bf Q},lnp}=\left( {W_{n}\kappa { } ^{3}_{n}\over \kappa { }
 ^{3}_{0}}\right) ^{1/2}\langle \tilde{\chi }_{0{\bf Q}}|
 \sum\limits^{3}_{l^{\prime\prime}=1}
 \hat{U}^{\text{eff}}_{l^{\prime\prime}}| \tilde{\psi
 }_{l0p_n}\rangle  \nonumber
\end{eqnarray}
[$\tilde{\chi }^{\left( 0\right) }_{0{\bf Q}}\equiv \tilde{\chi
 }_{0{\bf Q}}$  since $\tilde{\bm{\chi }}_{0{\bf Q}}$  has the
 three-atom component only,
see Eq.\ (\ref{chiv0Q})]. Thus, solutions of the effective
 single-channel
problem (\ref{EqpsiUeff}) provide all necessary information on
 multichannel
scattering.

Equation (\ref{EqpsiUeff}) describes a three-body problem with
separable two-body interactions. Following Lovelace
 \cite{Lovelace1964} (see
also books \cite{Schmid,Glockle} and an adaptation of this approach
 to 1D
problems in \cite{Melde2002}), the problem can be reduced to an
 integral
equation
\begin{equation}
X\left( q^\prime ,q_{0}\right) =2Z\left( q^\prime ,q_{0}\right) +{m{
 } ^{2}\over 2\kappa { } ^{3}_{0}}\int dq Z\left( q^\prime ,q\right)
 T_{1D}\left( k\left( q\right) \right) X\left( q,q_{0}\right)
 \label{Lovelace}
\end{equation}
for the symmetric transition amplitude
\begin{equation}
X\left( q^\prime ,q\right) =\sum\limits^{3}_{l=1}\langle \tilde{\chi
 }_{l^\prime 0q^\prime }|\sum\limits^{}_{l^{\prime\prime}\neq
 l^\prime }\hat{U}^{\text{eff}}_{l^{\prime\prime}}|\tilde{\psi
 }_{l0q}\rangle  \label{Xppp}
\end{equation}
(it is independent of $l^\prime $, see
 \cite{Lovelace1964,Schmid,Glockle}). Here
\begin{equation}
Z\left( q^\prime ,q\right) ={2\kappa { } ^{3}_{0}\over \pi m}{1\over
 mE+i0-q^{2}-qq^\prime -q^{\prime 2}}
\end{equation}
is the Green function for three free atoms, expressed in terms of the
momenta $q$ in two Jacobi coordinate sets, the two-body $T$ matrix
 $T_{1D}\left( k\right) $ is
given by Eq.\ (\ref{T1D}), and the relative momentum of two atoms is
expressed in terms of $q$ as
\begin{equation}
k\left( q\right) =\sqrt{mE+i0-3q^{2}/4} . \label{kq}
\end{equation}
\subsection{Transition probabilities}\label{TranProb}

Identifying the transmission and reflection amplitudes in Eq.\
(\ref{psias}) and using Eqs.\ (\ref{XUeff}) and (\ref{Xppp}) one can
express the reflection and transmission probabilities in terms of the
symmetric transition amplitude
\begin{eqnarray}
P^{\text{ref}}_{n^\prime n}={16\pi { } ^{2}\over 9}{m^{2}W_{n^\prime
 }W{ } _{n}\over p_{n^\prime }p{ } _{n}} {\kappa ^{3}_{n^\prime
 }\kappa { } ^{3}_{n}\over \kappa { } ^{6}_{0}}|X\left( -p_{n^\prime
 },p_{n}\right) |^{2} \nonumber
\\
P^{\text{tran}}_{n^\prime n}={16\pi { } ^{2}\over 9}{m^{2}W_{n^\prime
 }W{ } _{n}\over p_{n^\prime }p{ } _{n}} {\kappa ^{3}_{n^\prime
 }\kappa { } ^{3}_{n}\over \kappa { } ^{6}_{0}}|X\left( p_{n^\prime
 },p_{n}\right) |^{2}\qquad \left( n^\prime \neq n\right)  .
 \label{Preftr}
\\
P^{\text{tran}}_{nn}=\left|1-i{4\pi m\over 3p{ } _{n}}W_{n}\left(
 {\kappa { } _{n}\over \kappa { } _{0}}\right) ^{3}X\left(
 p_{n},p_{n}\right) \right|^{2} \nonumber
\end{eqnarray}
Atom-diatom collisions can also result in dissociation of the diatom.
This process is described by the last term in Eq.\ (\ref{psias}).
 However,
the total dissociation probability is related to probabilities of
 other
processes as
\begin{equation}
P^{\text{diss}}_{n}=1-\sum\limits^{}_{n^\prime }\left(
 P^{\text{tran}}_{n^\prime n}+P^{\text{ref}}_{n^\prime n}\right)  .
\end{equation}
Diatoms can be formed in three-atom collisions. The probability of
association per unit time can be found, like in the 3D case
\cite{Fedichev1996b}, from the equation
\begin{equation}
w^{\text{assoc}}_{n}\left( {\bf Q}\right) =2\pi \int {Ldp\over 2\pi }
|\langle {\bf Q}|\sum\limits^{3}_{l=1}\hat{\bm{U}}_{l}|n,p\rangle
|^{2}\delta \left( 3{p{ } ^{2}\over 4m}-{\kappa { } ^{2}_{n}\over
 m}-E\right)  . \label{wnass}
\end{equation}
Here
\begin{equation}
|{\bf Q}\rangle ={2\pi \over L}6^{-1/2}\sum\limits^{}_{\bm{\Pi
 }}\tilde{\bm{\chi }}_{0\bm{\Pi }({\bf Q})} \label{wfQbox}
\end{equation}
is the wavefunction of three free atoms with momenta $Q_{1}$,
 $Q_{2}$, and $Q_{3}$
($Q_{1}+Q_{2}+Q_{3}=0$), and the energy is $E={\bf Q}^{2}/\left(
 2m\right) $. The wavefunction (\ref{wfQbox})
is normalized per unit in the box of the length $L$ and symmetrized
 over all
6 possible permutations $\bm{\Pi }\left( {\bf Q}\right) $ of the
 momenta $Q_{i}$. The wavefunction of
the interacting atom and diatom in the state $n$,
\begin{equation}
|n,p\rangle =\left( {2\pi \over 3L}\right) ^{1
/2}\sum\limits^{3}_{l=1}\tilde{\bm{\psi }}_{lnp} \label{wfnpbox} ,
\end{equation}
has the box normalization too and is symmetrized over the atom
permutations.

In the system of $N_{\text{at}}$  atoms with momenta $Q_{i}$  the
 rate of formation of
molecules in the state $n$ can be evaluated by summation over all
 trios
\begin{equation}
{dN{ } _{n}\over dt}=\sum\limits^{}_{1\le i_1<i_2<i_3\le
 N_{\text{at}}}w^{\text{assoc}}_{n}\left( Q_{i_1}-{Q_{i_1}+Q_{i_2}+Q{
 } _{i_3}\over 3},Q_{i_2}-{Q_{i_1}+Q_{i_2}+Q{ } _{i_3}\over
 3},Q_{i_3}-{Q_{i_1}+Q_{i_2}+Q{ } _{i_3}\over 3}\right)  .
\end{equation}
Replacing the summation by integration with the atomic momentum
distribution $f_{\text{at}}\left( Q\right) $, one obtains the rate
 equation for the molecular
density $\rho _{n}=N_{n}/L$,
\begin{equation}
{d\rho { } _{n}\over dt}=\rho ^{3}_{\text{at}}\int d^{3}QK_{3n}\left(
 Q_{1}-{Q_{1}+Q_{2}+Q{ } _{3}\over 3},Q_{2}-{Q_{1}+Q_{2}+Q{ }
 _{3}\over 3},Q_{3}-{Q_{1}+Q_{2}+Q{ } _{3}\over 3}\right)
 f_{\text{at}}\left( Q_{1}\right) f_{\text{at}}\left( Q_{2}\right)
 f_{\text{at}}\left( Q_{3}\right)  ,
\end{equation}
where $\rho _{\text{at}}=N_{\text{at}}/L$ is the atomic density and
\begin{eqnarray}
K_{3n}\left( Q_{1},Q_{2},Q_{3}\right) ={L{ } ^{2}\over 3
!}w^{\text{assoc}}_{n}\left( Q_{1},Q_{2},Q_{3}\right)  \nonumber
\\
={2\pi ^{2}m^{3}\kappa { } ^{3}_{n}\over 9\kappa ^{6}_{0}p{ }
 _{n}}W_{n}\biggl\lbrack |\sum\limits^{3}_{j=1}T_{1D}\left( k\left(
 Q_{j}\right) \right) X\left( Q_{j},p_{n}\right) |^{2}+
|\sum\limits^{3}_{j=1}T_{1D}\left( k\left( Q_{j}\right) \right)
 X\left( Q_{j},-p_{n}\right) |^{2}\biggr\rbrack  \label{K3n}
\end{eqnarray}
is the association rate coefficient. [Here $p_{n}$  is given by Eq.\
(\ref{pn}) and $Q_{1}+Q_{2}+Q_{3}=0$ again.] This expression is
 derived using Eqs.\
(\ref{wnass}), (\ref{wfQbox}), (\ref{wfnpbox}), (\ref{X0Qmchan}),
(\ref{XUeff}), and the general relation between bound-free and
 bound-bound
transition amplitudes (see \cite{Schmid,Glockle}).
\begin{equation}
\langle \tilde{\chi }_{0{\bf Q}}{\bf |}
 \sum\limits^{3}_{l^{\prime\prime}=1}
 \hat{U}^{\text{eff}}_{l^{\prime\prime}}| \tilde{\psi
 }_{l0p_n}\rangle  = -{m\over 2}\left( 2\pi \kappa ^{3}_{0}\right)
 ^{-1/2}\sum\limits^{3}_{j=1}T_{1D}\left( k\left( Q_{j}\right) \right
)  \langle \tilde{\chi }_{j0Q_j}|
 \sum\limits^{}_{l^{\prime\prime}\neq
 j}\hat{U}^{\text{eff}}_{l^{\prime\prime}}| \tilde{\psi
 }_{l0p_n}\rangle
\end{equation}
The rate coefficient (\ref{K3n}) is defined according to conventional
chemical notation and is three times larger then the $K_{3}$  used in
\cite{Yurovsky2006,Yurovsky2008b}

\subsection{Special cases}\label{PrivCase}

\subsubsection{Two-channel case}

In this case \cite{Yurovsky2006}, when only one molecular channel
($m=c$)
is present, the vector $G_{m}\left( E\right) $ contains only one
 element,
\begin{equation}
G_{c}\left( E\right) ={g{ } _{c}\over E+i0-D{ } _{c}} . \label{Gm1}
\end{equation}
Therefore, the effective interaction strength is
\begin{equation}
U_{\text{eff}}\left( E\right) =U_{a}+{2|g_{c}|{ } ^{2}\over E+i0-D{ }
 _{c}} \label{Ueff2c}
\end{equation}
and Eq.\ (\ref{kappanpol}) for the poles of $T_{1D}$  is reduced to
 the
cubic equation \cite{Kheruntsyan1998,Yurovsky2005}
\begin{equation}
\kappa ^{3}+{m\over 2}U_{a}\kappa ^{2}+mD_{c}\kappa +{1\over
 2}m^{2}D_{c}U_{a}-m^{2}|g_{c}|^{2}=0 . \label{kappa2c}
\end{equation}
It has two positive roots $\kappa _{0,1}$  whenever  $U_{a}<0$ and
 $D_{c}<2|g_{c}|^{2}/U_{a}$,  and one
positive root $\kappa _{0}$  otherwise. The open-channel
 contributions to the
corresponding bound states are \cite{Yurovsky2006b},
\begin{equation}
W_{n}={\kappa ^{2}_{n}+mD{ } _{c}\over 3\kappa ^{2}_{n}+mU_{a}\kappa
 _{n}+mD{ } _{c}} .
\end{equation}
\subsubsection{Deactivation case}

The general problem with multiple molecular channels can be
substantially simplified in the case when one of the molecular
 channels,
the closed channel ($m=c$), is coupled to all other channels, while
 other
molecular channels, the deactivation products ($m\in \{d\}$), are not
 mutually
coupled ($d_{dd^\prime }=0$, $d\neq d^\prime $). In this case the
 effective interaction strength
can be expressed as
\begin{equation}
U_{\text{eff}}\left( E\right) =U_{a}+2\sum\limits^{}_{d}{|g_{d}|{ }
 ^{2}\over E+i0-D{ } _{d}}+2\left( |g_{c}|^{2}+g^{
*}_{c}\sum\limits^{}_{d}{d_{cd}g{ } _{d}\over E+i0-D{ } _{d}}\right)
 \left( E+i0-D_{c}-\sum\limits^{}_{d}{|d_{cd}|{ } ^{2}\over E+i0-D{ }
 _{d}}\right) ^{-1} .
\end{equation}
If the deactivation product states lie far below the open channel
threshold, i. e. $D_{d}<0$ and
\begin{equation}
|D_{d}|\gg \max(|d^{2}_{cd}/D_{c}|, |g_{d}d_{cd}/g_{c}|, |g^{2}_{d}
/U_{a}|, |E|, |D_{c}|, mU^{2}_{a}) ,
\end{equation}
the effective interaction strength is approximately the same as
in the two-channel case (\ref{Ueff2c}), and Eq.\ (\ref{kappan}) has
roots which are approximately determined by the same cubic equation
(\ref{kappa2c}).

Equation (\ref{kappan}) also has roots
\begin{equation}
\kappa _{d}\approx \sqrt{m|D_{d}|} , \label{kappad}
\end{equation}
one per each deactivation product state. The open-channel
contributions to the corresponding bound states are
\begin{equation}
W_{d}\approx {1\over 2}\sqrt{m}|g_{d}|^{2}|D_{d}|^{-3/2} . \label{Wd}
\end{equation}
The rate coefficient for deactivation of the diatom state $n$ onto all
states $\{d\}$ can be expressed, using Eqs.\ (\ref{Preftr}),
(\ref{Wd}) and
(\ref{Xpdpn}), as
\begin{equation}
K_{2n}={3p{ } _{n}\over 2m}\sum\limits^{}_{\{d\}}\left(
 P^{\text{ref}}_{dn}+P^{\text{tran}}_{dn}\right)
 =\sum\limits^{}_{\{d\}}|\gamma _{ad}\psi ^{\left( 0\right)
 }_{np_n}\left( 0,0,0\right) +\gamma _{cd}\psi ^{\left( c\right)
 }_{np_n}\left( 0,0\right) |^{2} , \label{K2n}
\end{equation}
where the coefficients $\gamma _{ad}$  and $\gamma _{cd}$  are
 expressed as
\begin{equation}
\gamma _{ad}=\left( {3^{5}4m\over |D_{d}|{ } ^{5}}\right) ^{1/4}
|g_{d}|U_{a}, \gamma _{cd}=2\left( {3^{3}m\over |D_{d}|{ }
 ^{5}}\right) ^{1/4}|g_{d}|g^{*}_{c}
\end{equation}
and the symmetric wavefunctions
\begin{eqnarray}
\psi ^{\left( 0\right) }_{np_n}\left( 0,0,0\right)
 ={\sqrt{3W_{n}\kappa { } _{n}}\over 2\pi } \left\lbrack 1
+{im^{2}\kappa { } _{n}\over 4\kappa { } ^{3}_{0}}\int dq
 {T_{1D}\left( k\left( q\right) \right) X\left( q,p_{n}\right) \over
 k\left( q\right) }\right\rbrack  \nonumber
\\
\label{psinpno}
\\
\psi ^{\left( c\right) }_{np_n}\left( 0,0\right) =-{mg{ } _{c}\over
 \pi }\sqrt{{W_{n}\kappa { } _{n}\over 2}}\biggl\lbrack {1\over
 \kappa ^{2}_{n}+mD{ } _{c}}+{m\kappa { } _{n}\over 2\kappa { }
 ^{3}_{0}}\int dq{T_{1D}\left( k\left( q\right) \right) X\left(
 q,p_{n}\right) \over U_{a}\left( k^{2}\left( q\right) -mD_{c}\right)
 +2m|g_{c}|{ } ^{2}}\biggr\rbrack  . \nonumber
\end{eqnarray}
are calculated with the symmetric transition amplitudes $X\left(
 q,p_{n}\right) $ and the
two-body $T$ matrix $T_{1D}\left( k\right) $ for the two-channel case
 (see App.\ \ref{AppDeac}).

\section{Regularization of one-dimensional Faddeev-Lovelace
equations}\label{SolFaddLov}

The previous section demonstrates that a solution of Eq.\
(\ref{Lovelace}) provides all necessary information for evaluation of
transition probabilities. This equation is a particular case of the
 general
Lovelace equation (see \cite{Lovelace1964,Schmid,Glockle,Melde2002})
 for
separable potentials $V\tilde{v}^{*}\left( q\right) \tilde{v}\left(
 q^\prime \right) $ with the interaction strength $V$ and
formfactors $\tilde{v}\left( q^\prime \right) $. For zero-range
 potentials the formfactors are constant
functions, $\tilde{v}\left( q\right) \equiv \left( 2\pi \right) ^{-1
/2}$, in the momentum representation and $v\left( z\right) =\delta
 \left( z\right) $ in
the coordinate representation [indeed, $Vv^{*}\left( z\right) \int
 dz^\prime v\left( z^\prime \right) \varphi \left( z^\prime \right)
 =V\delta \left( z\right) \varphi \left( z\right) $ in
this case]. In the general case the two-body $T$ matrix is expressed
 as $\tilde{v}^{*}T_{1D}\tilde{v}$
in terms of the function $T_{1D}$  in Eq.\ (\ref{Lovelace}) and the
 free Green
function $Z\left( q^\prime ,q\right) $ has the form
\begin{equation}
Z\left( q^\prime ,q\right) ={4\kappa { } ^{3}_{0}\over m}
 {\tilde{v}^{*}\left( q+q^\prime /2\right) \tilde{v}\left( -q
/2-q^\prime \right) \over mE+i0-q^{2}-qq^\prime -q^{\prime 2}} .
\end{equation}
As in the 3D case (see \cite{Schmid}), solution of 1D Lovelace
equations can be troubled by singularities. The regularization
 procedure
described below is applicable to the general case of separable
 potentials.

The singularities can arise from both the two-body $T$ matrix and the
free Green function. Poles of $T_{1D}\left( k\left( q\right) \right)
 $ at $q=\pm p_{n}$  are related to two-body
bound states and appear for each scattering process, whenever the
 energy
lies above the lowest bound state, $E>-\max\left( \kappa
 ^{2}_{n}\right) /m$ [see Eq.\ (\ref{pn})]. The
free Green function $Z\left( q^\prime ,q\right) $ has singularities
 at $q=q_{\pm }\left( q^\prime \right) =-q^\prime /2\pm k\left(
 q^\prime \right) $,
where the function $k\left( q\right) $ is given by Eq.\ (\ref{kq}).
 Since $q^\prime $  and $k\left( q^\prime \right) $
are the momenta $q$ and $k$ in one of the Jacobi coordinate systems,
 Eq.\
(\ref{JacRel}) allows to recognize $q_{\pm }\left( q^\prime \right) $
 as the momenta $q$ in other
systems, where the atoms are permuted. Therefore the poles correspond
 to
non-diffractive collisions where the momenta keep their values and
 are only
exchanged by the atoms. Then the transformations $q_{\pm }\left(
 q_{\pm }\left( q^\prime \right) \right) $ can lead only
to one of the three momenta $q^\prime $, $q_{+}\left( q^\prime \right
) $, and $q_{-}\left( q^\prime \right) $. This property will be
used below (see Tab.\ \ref{Tablezpm}). Since they lie outside of the
 real
axis for $q^\prime =\pm p_{n}$, the singularities of $Z\left(
 q^\prime ,q\right) $ play a role only in the case
when $q^\prime $  corresponds to three free atoms. In the source term
 $Z\left( q^\prime ,q_{0}\right) $ of
Eq.\ (\ref{Lovelace}) the singularities can appear only when both
 $q^\prime $   and
$q_{0}$  correspond to three free atoms. However, the present work
 does not deal
with this case of free-free transitions. Unlike the 3D case, where a
partial wave expansion leads to logarithmic singularities of $Z\left(
 q^\prime ,q\right) $, in
the present 1D case they are simple poles.

For further derivation let us introduce dimensionless variables.
Convenient energy and momentum scales are based on $g$, a
 characteristic value
of the channel coupling strengths $g_{m}$. The dimensionless momentum
 and energy
are expressed as
\begin{equation}
\zeta ={\sqrt{3}\over 2}\left( mg\right) ^{-2/3}q ,\qquad \epsilon
 =m^{-1/3}g^{-4/3}E .
\end{equation}
An introduction of new functions in place of the transition amplitude,
the free Green function, and the two-body $T$ matrix,
\begin{eqnarray}
F\left( \zeta ,\zeta _{0}\right) =-{\pi \over \sqrt{3}}m^{7/3}g^{4
/3}\kappa ^{-3}_{0}\sqrt{\epsilon +i0-\zeta { } ^{2}} X\left(
 q,q_{0}\right)  \label{FX}
\\
S\left( \zeta ^\prime ,\zeta \right) =-{2\pi \over \sqrt{3}}m^{7
/3}g^{4/3}\kappa ^{-3}_{0}\sqrt{\epsilon +i0-\zeta ^{\prime 2}}Z\left
( q^\prime ,q\right) ={1\over \zeta -\zeta _{+}\left( \zeta ^\prime
 \right) }-{1\over \zeta -\zeta _{-}\left( \zeta ^\prime \right) }
 \label{Szpz}
\\
r\left( \zeta \right) =-{m{ } ^{1/3}\over 2\pi g{ } ^{2/3}}\left(
 \epsilon +i0-\zeta ^{2}\right) ^{-1/2}T_{1D}\left( k\left( q\right)
 \right) =-{1\over 2\pi }\left(
 \tilde{U}^{-1}_{\text{eff}}\sqrt{\epsilon +i0-\zeta { } ^{2}}
+{i\over 2}\right) ^{-1} , \label{rz}
\end{eqnarray}
allows to exclude the square-root singularity from $X\left(
 q,q_{0}\right) $. Here
\begin{equation}
\zeta _{\pm }\left( \zeta \right) =-{1\over 2}\zeta \pm
 {\sqrt{3}\over 2}\sqrt{\epsilon +i0-\zeta { } ^{2}} ,\qquad
 \tilde{U}_{\text{eff}}=m^{1/3}g^{-2/3}U_{\text{eff}} .
\end{equation}
Equation (\ref{Lovelace}) is transformed then to the form
\begin{equation}
F\left( \zeta ^\prime ,\zeta _{0}\right) =S\left( \zeta ^\prime
 ,\zeta _{0}\right) +\int\limits^{\infty }_{-\infty }d\zeta  S\left(
 \zeta ^\prime ,\zeta \right) r\left( \zeta \right) F\left( \zeta
 ,\zeta _{0}\right)  .
\end{equation}
\subsection{Singularities of the free Green function}

According to Eq.\ (\ref{Szpz}) $S\left( \zeta ^\prime ,\zeta \right)
 $ has simple poles on the real
axis whenever $\zeta ^{\prime 2}\le \epsilon $. The pole
 contributions can be separated using
the symbolic expression
\begin{equation}
{1\over \zeta -\zeta _{0}-i0}=\text{v.p.}{1\over \zeta -\zeta { }
 _{0}}+i\pi \delta \left( \zeta -\zeta _{0}\right)  , \label{singfrac}
\end{equation}
leading to the integral equation
\begin{equation}
F\left( \zeta ^\prime ,\zeta _{0}\right) -i\pi r\left( \zeta _{
+}\right) F\left( \zeta _{+},\zeta _{0}\right) -i\pi r\left( \zeta
 _{-}\right) F\left( \zeta _{-},\zeta _{0}\right) ={1\over \zeta
 _{0}-\zeta { } _{+}}-{1\over \zeta _{0}-\zeta { } _{-}}
+\int\limits^{\infty }_{-\infty }d\zeta \text{ v.p.}\left( {1\over
 \zeta -\zeta { } _{+}}-{1\over \zeta -\zeta { } _{-}}\right) r\left(
 \zeta \right) F\left( \zeta ,\zeta _{0}\right)  . \label{Fzpzpzm}
\end{equation}
Here $\zeta _{\pm }=\zeta _{\pm }\left( \zeta ^\prime \right) $,
 symbol v.p. means Cauchy principal value integral in
the vicinity of $\zeta _{\pm }$, and the source term is non-singular
 for the problems
considered here, as it was mentioned above. The following derivation is
 based
on the algebraic properties of the transformation $\zeta _{\pm }\left
( \zeta \right) $. A direct
calculation of products of these transformations (see Tab.\
 \ref{Tablezpm})
demonstrates that $\zeta _{\pm }\left( \zeta _{\pm }\right) $ leads
 again to one of the three values $\zeta ^\prime $, $\zeta _{+}$, or
$\zeta _{-}$. Therefore, replacement $\zeta ^\prime $  by $\zeta _{
+}$  or by $\zeta _{-}$  in Eq.\ (\ref{Fzpzpzm})
leads to two additional linear equations in $F\left( \zeta ^\prime
 ,\zeta _{0}\right) $, $F\left( \zeta _{+},\zeta _{0}\right) $, and
$F\left( \zeta _{-},\zeta _{0}\right) $. Then they form a system of
 linear equations with the matrix
\begin{equation}
A=\left(\begin{array}{ccc}1&-i\pi r\left( \zeta _{+}\right) &-i\pi
 r\left( \zeta _{-}\right)  \\
 -i\pi r\left( \zeta ^\prime \right) &1&-i\pi r\left( \zeta
 _{-}\right)  \\
 -i\pi r\left( \zeta ^\prime \right) &-i\pi r\left( \zeta _{+}\right)
 &1\end{array}\right)  . \label{matrA}
\end{equation}
\begin{table}

\caption{Algebraic properties of
transformations $\zeta _{\pm }$\label{Tablezpm}}
\begin{ruledtabular} \begin{tabular}{ccccc}
&$\zeta _{-}\left( \zeta _{-}\left( \zeta ^\prime \right) \right)
 $&$\zeta _{+}\left( \zeta _{-}\left( \zeta ^\prime \right) \right)
 $&$\zeta _{-}\left( \zeta _{+}\left( \zeta ^\prime \right) \right)
 $&$\zeta _{+}\left( \zeta _{+}\left( \zeta ^\prime \right) \right)
 $\\
\hline $\zeta ^\prime <-\sqrt{\epsilon }/2$&$\zeta ^\prime $&$\zeta
 _{+}\left( \zeta ^\prime \right) $&$\zeta ^\prime $&$\zeta _{-}\left
( \zeta ^\prime \right) $\\
$-\sqrt{\epsilon }/2<\zeta ^\prime <\sqrt{\epsilon }/2$&$\zeta
 ^\prime $&$\zeta _{+}\left( \zeta ^\prime \right) $&$\zeta _{-}\left
( \zeta ^\prime \right) $&$\zeta ^\prime $\\
$\zeta ^\prime >\sqrt{\epsilon }/2$&$\zeta _{+}\left( \zeta ^\prime
 \right) $&$\zeta ^\prime $&$\zeta _{-}\left( \zeta ^\prime \right)
 $&$\zeta ^\prime $\\

\end{tabular}\end{ruledtabular}

\end{table} Solution of this system leads to the integral equation in
$F\left( \zeta ^\prime ,\zeta _{0}\right) $
\begin{equation}
F\left( \zeta ^\prime ,\zeta _{0}\right) =\bar{S}\left( \zeta ^\prime
 ,\zeta _{0}\right) +\int\limits^{\infty }_{-\infty }d\zeta \text{
 v.p.}\bar{S}\left( \zeta ^\prime ,\zeta \right) r\left( \zeta \right
) F\left( \zeta ,\zeta _{0}\right)  .
\end{equation}
Here
\begin{equation}
\bar{S}\left( \zeta ^\prime ,\zeta \right) =\cases{{s\over \zeta
 -\zeta ^\prime }+{s{ } _{+}\over \zeta -\zeta { } _{+}}+{s{ }
 _{-}\over \zeta -\zeta { } _{-}} & $\zeta ^\prime \le \sqrt{\epsilon
 }$ \cr S\left( \zeta ^\prime ,\zeta \right)  & $\zeta ^\prime
>\sqrt{\epsilon }$ }
\end{equation}
and the parameters $s$, $s_{+}$, and $s_{-}$  are expressed as
\begin{eqnarray}
s=\left( A^{-1}\right) _{12}\text{sign}\left( \zeta ^\prime
+\sqrt{\epsilon }/2\right) +\left( A^{-1}\right)
 _{13}\text{sign}\left( \zeta ^\prime -\sqrt{\epsilon }/2\right)
 \nonumber
\\
s_{+}=\left( A^{-1}\right) _{11}-\left( A^{-1}\right)
 _{13}\text{sign}\left( \zeta ^\prime -\sqrt{\epsilon }/2\right)
\\
s_{-}=-\left( A^{-1}\right) _{11}-\left( A^{-1}\right)
 _{12}\text{sign}\left( \zeta ^\prime +\sqrt{\epsilon }/2\right)
 \nonumber
\end{eqnarray}
in terms of the inverted matrix (\ref{matrA}).

\subsection{Singularities of two-body $T$ matrix}

Following derivation is a generalization of the approach
\cite{Dodd1972} to the case of multiple bound states. Since $r\left(
 \zeta \right) $ is an even
function of $\zeta $  [see Eq.\ (\ref{rz})] and $\bar{S}\left( -\zeta
 ^\prime ,\zeta \right) =\bar{S}\left( \zeta ^\prime ,-\zeta \right)
 $ (this property
can be directly proven), the equations for the odd and even
 components,
\begin{equation}
F_{\pm }\left( \zeta ,\zeta _{0}\right) =F\left( \zeta ,\zeta
 _{0}\right) \pm F\left( -\zeta ,\zeta _{0}\right)  , \label{Fpm}
\end{equation}
can be separated as
\begin{equation}
F_{\pm }\left( \zeta ^\prime ,\zeta _{0}\right) =\bar{S}_{\pm }\left(
 \zeta ^\prime ,\zeta _{0}\right) +\int\limits^{\infty }_{0}d\zeta
 \text{ v.p.}\bar{S}_{\pm }\left( \zeta ^\prime ,\zeta \right) r\left
( \zeta \right) F_{\pm }\left( \zeta ,\zeta _{0}\right) { } _{,}
 \label{LovSym}
\end{equation}
where
\begin{equation}
\bar{S}_{\pm }\left( \zeta ^\prime ,\zeta \right) =\bar{S}\left(
 \zeta ^\prime ,\zeta \right) \pm \bar{S}\left( -\zeta ^\prime ,\zeta
 \right)  .
\end{equation}
The odd and even component in 1D scattering are analogs of 3D partial
waves \cite{Lipkin}.

Equations (\ref{LovSym}) still contain singularities related to poles
of $r\left( \zeta \right) $ at $=\zeta _{n}+i0$, where $\zeta
 _{n}={\sqrt{3}\over 2}\left( mg\right) ^{-2/3}p_{n}$. In the pole
 vicinity the function
$r\left( \zeta \right) $ has the limiting behavior
\begin{equation}
r\left( \zeta \right) \mathrel{ \mathop \sim  _{\zeta \rightarrow
 \zeta _n}}{-ir{ } _{n}\over \zeta -\zeta _{n}-i0} , r_{n}=i\text{
 Res}\left( r\left( \zeta \right) ,\zeta =\zeta _{n}\right) ={1\over
 \pi }{\kappa { } ^{2}_{n}\over m^{4/3}g{ } ^{4/3}}{W{ } _{n}\over
 \zeta { } _{n}}
\end{equation}
[see Eqs.\ (\ref{resT1D}) and (\ref{rz})]. The contributions of the
singularities can be separated  using Eq.\ (\ref{singfrac}) leading
 to the
regularized equations
\begin{equation}
F_{\pm }\left( \zeta ^\prime ,\zeta _{0}\right) =\bar{S}_{\pm }\left(
 \zeta ^\prime ,\zeta _{0}\right) +\pi
 \sum\limits^{}_{n}r_{n}\bar{S}_{\pm }\left( \zeta ^\prime ,\zeta
 _{n}\right) F_{\pm }\left( \zeta _{n},\zeta _{0}\right) +\text{v.p.
 }\int\limits^{\infty }_{0}d\zeta  \bar{S}_{\pm }\left( \zeta ^\prime
 ,\zeta \right) r\left( \zeta \right) F_{\pm }\left( \zeta ,\zeta
 _{0}\right) { } _{,}
\end{equation}
where the Cauchy principal value integral is taken for all
singularities in $\bar{S}_{\pm }\left( \zeta ^\prime ,\zeta \right) $
 and $r\left( \zeta \right) $.

Source terms in these equations have forms of linear combinations of
the functions $\bar{S}_{\pm }\left( \zeta ^\prime ,\zeta _{n}\right)
 $ (for the problems under consideration $\zeta _{0}$  is equal
to one of $\zeta _{n}$). Thus the solution can be expressed as a
 linear combination
\begin{equation}
F_{\pm }\left( \zeta ^\prime ,\zeta _{0}\right) =\bar{F}_{\pm }\left(
 \zeta ^\prime ,\zeta _{0}\right) +\pi \sum\limits^{}_{n}r_{n}F_{\pm
 }\left( \zeta _{n},\zeta _{0}\right) \bar{F}_{\pm }\left( \zeta
 ^\prime ,\zeta _{n}\right)  \label{FpmFpmbar}
\end{equation}
of solutions $\bar{F}_{\pm }\left( \zeta ,\zeta _{0}\right) $ of
 integral equations
\begin{equation}
\bar{F}_{\pm }\left( \zeta ^\prime ,\zeta _{n}\right) =\bar{S}_{\pm
 }\left( \zeta ^\prime ,\zeta _{n}\right) +\text{v.p.
 }\int\limits^{\infty }_{0}d\zeta  \bar{S}_{\pm }\left( \zeta ^\prime
 ,\zeta \right) r\left( \zeta \right) \bar{F}_{\pm }\left( \zeta
 ,\zeta _{n}\right) { } _{,} \label{Fbarpm}
\end{equation}
where the source terms do not contain unknown functions. Solving of
these equations also allows the evaluation of the values of $F_{\pm
 }$  at specific
points, $F_{\pm }\left( \zeta _{n},\zeta _{0}\right) $, involved into
 Eq.\ (\ref{FpmFpmbar}). They are
determined by the system of linear algebraic equations
\begin{equation}
F_{\pm }\left( \zeta _{n},\zeta _{0}\right) -\pi
 \sum\limits^{}_{n^\prime }r_{n^\prime }\bar{F}_{\pm }\left( \zeta
 _{n},\zeta _{n^\prime }\right) F_{\pm }\left( \zeta _{n^\prime
 },\zeta _{0}\right) =\bar{F}_{\pm }\left( \zeta _{n},\zeta
 _{0}\right)  , \label{Fpmznz0}
\end{equation}
which are obtained by setting $\zeta ^\prime =\zeta _{n}$  in Eqs.\
(\ref{FpmFpmbar}).

\section*{Conclusions}

Multichannel resonant interactions can be described by a model
involving one atomic and several molecular states. Even in the case of
zero-range interactions this model predicts several two-body bound
 states
(diatoms), which are superpositions of two-atomic and molecular
 components.
Orthogonality of the diatom wavefunctions is proven.

In three-body problems the multichannel behavior leads both to an
effective energy-dependent interactions and multichannel asymptotic
wavefunctions. The last effect leads to rescaling of the
 probabilities of
elastic and inelastic atom-diatom collisions and of three-atom
 association.
All the probabilities can be expressed in terms of the symmetric
 transition
amplitude for an effective single-channel problem. This amplitude can
 be
evaluated as a solution of the Faddeev-Lovelace equation.
 Singularities in
this equation can be regularized using their algebraic properties. The
regularization procedure is applicable to generic 1D three-body
 problems
with separable interactions.

The algorithm of solution of a resonant 1D three-body problem consists
then of the following steps:

--- functions $\bar{F}_{\pm }\left( \zeta ,\zeta _{n}\right) $ are
 determined as solutions of the regularized
integral equations (\ref{Fbarpm}) for each bound state $n$;

--- the system of linear equations (\ref{Fpmznz0}) is solved for
$F_{\pm }\left( \zeta _{n},\zeta _{0}\right) $;

--- $F_{\pm }\left( \zeta ,\zeta _{0}\right) $ is calculated by using
 of Eq.\ (\ref{FpmFpmbar});

--- $X\left( q,q_{0}\right) $ is determined using Eqs.\ (\ref{FX})
 and (\ref{Fpm});

--- transition probabilities are evaluated using Eqs.\ (\ref{Preftr})
(\ref{K3n}), (\ref{K2n}) and (\ref{psinpno}).

\appendix

\section{Orthogonality of the diatom states}\label{AppOrt}

Integration of Eqs.\ (\ref{cchan20}) and (\ref{phi0Ueff}) over an
infinitesimal interval including $z=0$ allows to represent them in
 the forms
\begin{eqnarray}
{1\over m}\hat{\partial }\varphi ^{\left( 0\right) }=U_{a}\varphi
 ^{\left( 0\right) }\left( 0\right) +\sqrt{2}\sum\limits^{}_{m}g^{
*}_{m}\varphi ^{\left( m\right) } \label{cchan20d}
\\
{1\over m}\hat{\partial }\varphi ^{\left( 0\right)
 }_{n}=U_{\text{eff}}\left( E_{n}\right) \varphi ^{\left( 0\right)
 }_{n}\left( 0\right)  , \label{phi0Ueffd}
\end{eqnarray}
where the wavefunction derivative discontinuity operator
 $\hat{\partial }$ acts as
\begin{equation}
\hat{\partial }\varphi ={\partial \varphi \over \partial z}|_{z=0
+0}-{\partial \varphi \over \partial z}|_{z=0-0} .
\end{equation}
Equations  (\ref{cchan20d}) and  (\ref{cchan2m})  allow to prove the
identity relation involving wavefunctions of two diatom states, $n$
 and $n^\prime $,
\begin{equation}
\varphi _{n^\prime }^{(0)*}\left( 0\right) {1\over m}\hat{\partial
 }\varphi ^{\left( 0\right) }_{n}+E_{n}\sum\limits^{}_{m}\varphi
 ^{\left( m\right) *}_{n^\prime }\varphi ^{\left( m\right)
 }_{n}=\varphi ^{\left( 0\right) }_{n}\left( 0\right) {1\over
 m}\hat{\partial }\varphi _{n^\prime }^{(0)*}+E_{n^\prime
 }\sum\limits^{}_{m}\varphi ^{\left( m\right) *}_{n^\prime }\varphi
 ^{\left( m\right) }_{n} . \label{idrelnnp}
\end{equation}
Multiplication of Eq.\ (\ref{phi0Ueff})  for $\varphi ^{\left(
 0\right) }_{n}$  by $\varphi _{n^\prime }^{(0)*}$   with
following subtraction of the complex conjugate of the similar product
 with
exchanged $n$ and $n^\prime $  leads to the following expression
\begin{eqnarray}
\left( E_{n}-E_{n^\prime }\right) \varphi _{n^\prime }^{(0)*}\left(
 z\right) \varphi ^{\left( 0\right) }_{n}\left( z\right) =-{1\over
 m}{\partial \over \partial z}\left\lbrack \varphi _{n^\prime }^{(0)
*}\left( z\right) {\partial \over \partial z}\varphi ^{\left( 0\right
) }_{n}\left( z\right) -\varphi ^{\left( 0\right) }_{n}\left( z\right
) {\partial \over \partial z}\varphi _{n^\prime }^{(0)*}\left(
 z\right) \right\rbrack  \nonumber
\\
+\delta \left( z\right) \left\lbrack U_{\text{eff}}\left( E_{n}\right
) -U_{\text{eff}}\left( E_{n^\prime }\right) \right\rbrack \varphi
 _{n^\prime }^{(0)*}\left( 0\right) \varphi ^{\left( 0\right)
 }_{n}\left( 0\right)
\end{eqnarray}
Integrating this expression from $-\infty $ to $\infty $ and using
 Eqs.\
(\ref{idrelnnp}) and (\ref{phi0Ueffd}) one gets the equation
\begin{equation}
\left( E_{n}-E_{n^\prime }\right) \left\lbrack  \int\limits^{\infty
 }_{-\infty }dz\varphi _{n^\prime }^{(0)*}\left( z\right) \varphi
 ^{\left( 0\right) }_{n}\left( z\right) +\sum\limits^{}_{m}\varphi
 ^{\left( m\right) }_{n^\prime }\varphi ^{\left( m\right)
 }_{n}\right\rbrack =0 .
\end{equation}
It leads to the orthogonality conditions (\ref{DiatomOrt}) for
non-degenerate states, while the degenerate states can be always
orthogonalized by a simple linear transformation.

\section{Residues of the two-body $T$ matrix}\label{AppResT}

For scattering on a generic potential $\hat{U}$, the two-body $T$
 matrix, as a
function of the collision energy $E$, has poles at the bound state
 energies
$E_{n}$  with the residues
\begin{equation}
\text{Res}\left( T_{1D},E=E_{n}\right) =\hat{U}|\varphi _{n}\rangle
 \langle \varphi _{n}|\hat{U} .
\end{equation}
In the present case $\hat{U}=U_{\text{eff}}\left( E\right) \delta
 \left( z\right) $ and the bound state wavefunction is
given by Eq.\ (\ref{phin0}). Taking into account Eqs.\ (\ref{kappan})
 and
(\ref{Wn}) one obtains
\begin{equation}
\text{Res}\left( T_{1D}\left( k\right) ,k=i\kappa _{n}\right)
 =-2i{\kappa { } ^{2}_{n}\over m}W_{n} . \label{resT1D}
\end{equation}
\section{Relations between transition amplitudes in the two-channel
and deactivation cases}\label{AppDeac}

The three-body asymptotic functions $\tilde{\chi }_{ldp_d}\left(
 q_{l},k_{l}\right) $, corresponding to
the deactivation product channels, are obtained by substitution of
 Eqs.\
(\ref{kappad}) and (\ref{pn}) into Eq.\ (\ref{chilnp}). For large $
|D_{d}|$ they
are approximately independent of the momentum $k_{l}$,
\begin{equation}
\tilde{\chi }_{ldp_d}\left( q_{l},k_{l}\right) \approx \left( {2\over
 \pi \kappa { } _{d}}\right) ^{1/2}\delta \left( q_{l}-p_{d}\right)
\end{equation}
(this means that sizes of the deactivation-product bound states are
negligibly small compared to other relevant scales). The transition
amplitude (\ref{Xmchan}) can be then transformed, using Eqs.\
(\ref{Umchan}) and (\ref{chi0chi}), to the form
\begin{equation}
X_{l^\prime dp_d,lnp_n}\approx {1\over 2\pi }\left( {2W{ } _{d}\over
 \pi \kappa { } _{d}}\right) ^{1/2}\int dk_{l^\prime
 }\sum\limits^{}_{l^{\prime\prime}\neq l^\prime }\left\lbrack
 U_{a}\int dk_{l^{\prime\prime}}\tilde{\psi }^{\left( 0\right)
 }_{lnp_n}\left( q_{l^{\prime\prime}},k_{l^{\prime\prime}}\right)
+2\left( {\pi \over 3}\right) ^{1/2}g^{*}_{c}\tilde{\psi }^{\left(
 c\right) }_{lnp_n}\left( q_{l^{\prime\prime}}\right) \right\rbrack  ,
\end{equation}
where $n=0,1$ and $\tilde{\psi }^{\left( 0,c\right) }_{lnp_n}$  are
 solutions of the two-channel problem,
which does not involve deactivation product states. Changing the
integration variable $k_{l^\prime }$  by $q_{l^{\prime\prime}}$
 allows us to recognize in the integrals
the coordinate-representations of the open and closed channel
 wavefunctions
[see Eq.\ (\ref{psi3})] in the origin. Finally, the symmetric
 transition
amplitude (\ref{Xppp}) involving the deactivation product states can
 be
expressed using Eq.\ (\ref{XUeff}) as
\begin{equation}
X\left( \pm p_{d},p_{n}\right) =3\left( {3\over W_{n}\kappa { }
 ^{3}_{n}}\right) ^{1/2}{\kappa { } ^{3}_{0}\over \kappa { }
 ^{2}_{d}}\left\lbrack U_{a}\psi ^{\left( 0\right) }_{np_n}\left(
 0,0,0\right) +\sqrt{{2\over 3}}g^{*}_{c}\psi ^{\left( c\right)
 }_{np_n}\left( 0,0\right) \right\rbrack  , \label{Xpdpn}
\end{equation}
in terms of the values at the origin of the symmetric wavefunctions in
the two-channel case
\begin{eqnarray}
\psi ^{\left( 0\right) }_{np_n}\left( 0,0,0\right) =3^{-1
/2}\sum\limits^{3}_{l=1}\psi ^{\left( 0\right) }_{lnp_n}\left(
 0,0,0\right) =\left( {W{ } _{n}\over 3}\right) ^{1/2}\left( 2\pi
 \right) ^{-3/2}\sum\limits^{3}_{l=1}\int dq_{j}dk_{j}\tilde{\psi
 }_{lnp_n}\left( q_{j},k_{j}\right)  \nonumber
\\
\label{psiorigin}
\\
\psi ^{\left( c\right) }_{np_n}\left( 0,0\right) =3^{-1
/2}\sum\limits^{3}_{l=1}\psi ^{\left( c\right) }_{lnp_n}\left(
 0,0\right) =\left( {W{ } _{n}\over \pi }\right) ^{1/2}{g{ }
 _{c}\over 2\pi }\sum\limits^{3}_{l=1}\int {dq{ } _{j}\over E
+i0-3q^{2}_{j}/\left( 4m\right) }\int dk_{j}\tilde{\psi
 }_{lnp_n}\left( q_{j},k_{j}\right)  \nonumber
\end{eqnarray}
[see Eqs.\ (\ref{psi3}), (\ref{psimom}), and (\ref{Gm1})].

These values can be also expressed in terms of the symmetric
transition amplitude for the two-channel case in the following way.
Let us write out the Lippmann-Schwinger equation for the effective
single-channel problem
\begin{equation}
\tilde{\psi }_{lnp}\left( q_{j},k_{j}\right) =\tilde{\chi
 }_{lnp}\left( q_{j},k_{j}\right) \delta _{jl}+\left( E
+i0-\hat{H}^{00}_{0}-\hat{U}^{\text{eff}}_{j}\right)
 ^{-1}\sum\limits^{}_{l^\prime \neq j}\hat{U}^{\text{eff}}_{l^\prime
 }\left\lbrack \tilde{\psi }_{lnp}\right\rbrack  .
 \label{LippSchwUeff}
\end{equation}
Here the Green function for the free $j$ th atom and diatom is the
integral operator
\begin{equation}
\left( E+i0-\hat{H}^{00}_{0}-\hat{U}^{\text{eff}}_{j}\right)
 ^{-1}\left\lbrack \tilde{\psi }\right\rbrack =g_{0}\left(
 q_{j},k_{j}\right) \int dk^\prime _{j}\left\lbrack \delta \left(
 k^\prime _{j}-k_{j}\right) +{1\over 2\pi }T_{1D}\left( k\left(
 q_{j}\right) \right) g_{0}\left( q_{j},k^\prime _{j}\right)
 \right\rbrack \tilde{\psi }\left( q_{j},k^\prime _{j}\right)  ,
\end{equation}
expressed in terms of the two-body $T$ matrix (\ref{T1D}) and the
 Green
function for three free atoms,
\begin{equation}
g_{0}\left( q,k\right) =\left\lbrack E+i0-{3\over 4m}q^{2}-{1\over
 m}k^{2}\right\rbrack ^{-1} .
\end{equation}
The incident wave $\tilde{\chi }_{lnp}\left( q,k\right) $ [see Eq.\
(\ref{chilnp})] is proportional
to $g_{0}\left( q,k\right) $ (it is a general property of separable
 potentials, see
\cite{Schmid,Glockle}). Substitution of $\psi $ given by Eq.\
(\ref{LippSchwUeff})
into Eq.\ (\ref{psiorigin}) allows to express the integrals in Eq.\
(\ref{psiorigin}) in terms of the transition amplitudes for the
 effective
single-channel problem. Some algebra leads to the values of symmetric
wavefunctions (\ref{psinpno}).


\begin{thebibliography}{54}
\expandafter\ifx\csname natexlab\endcsname\relax\def\natexlab#1{#1}\fi
\expandafter\ifx\csname bibnamefont\endcsname\relax
  \def\bibnamefont#1{#1}\fi
\expandafter\ifx\csname bibfnamefont\endcsname\relax
  \def\bibfnamefont#1{#1}\fi
\expandafter\ifx\csname citenamefont\endcsname\relax
  \def\citenamefont#1{#1}\fi
\expandafter\ifx\csname url\endcsname\relax
  \def\url#1{\texttt{#1}}\fi
\expandafter\ifx\csname urlprefix\endcsname\relax\def\urlprefix{URL }\fi
\providecommand{\bibinfo}[2]{#2}
\providecommand{\eprint}[2][]{\url{#2}}

\bibitem[{\citenamefont{Greiner et~al.}(2001)\citenamefont{Greiner, Bloch,
  Mandel, H\"ansch, and Esslinger}}]{Greiner2001}
\bibinfo{author}{\bibfnamefont{M.}~\bibnamefont{Greiner}},
  \bibinfo{author}{\bibfnamefont{I.}~\bibnamefont{Bloch}},
  \bibinfo{author}{\bibfnamefont{O.}~\bibnamefont{Mandel}},
  \bibinfo{author}{\bibfnamefont{T.~W.} \bibnamefont{H\"ansch}},
  \bibnamefont{and}
  \bibinfo{author}{\bibfnamefont{T.}~\bibnamefont{Esslinger}},
  \bibinfo{journal}{Phys. Rev. Lett.} \textbf{\bibinfo{volume}{87}},
  \bibinfo{pages}{160405} (\bibinfo{year}{2001}).

\bibitem[{\citenamefont{Moritz et~al.}(2005)\citenamefont{Moritz, Stoferle,
  Gunter, Kohl, and Esslinger}}]{Moritz2005}
\bibinfo{author}{\bibfnamefont{H.}~\bibnamefont{Moritz}},
  \bibinfo{author}{\bibfnamefont{T.}~\bibnamefont{Stoferle}},
  \bibinfo{author}{\bibfnamefont{K.}~\bibnamefont{Gunter}},
  \bibinfo{author}{\bibfnamefont{M.}~\bibnamefont{Kohl}}, \bibnamefont{and}
  \bibinfo{author}{\bibfnamefont{T.}~\bibnamefont{Esslinger}},
  \bibinfo{journal}{Phys. Rev. Lett.} \textbf{\bibinfo{volume}{94}},
  \bibinfo{eid}{210401}  (\bibinfo{year}{2005}).

\bibitem[{\citenamefont{Kinoshita et~al.}(2004)\citenamefont{Kinoshita, Wenger,
  and Weiss}}]{Kinoshita2004}
\bibinfo{author}{\bibfnamefont{T.}~\bibnamefont{Kinoshita}},
  \bibinfo{author}{\bibfnamefont{T.}~\bibnamefont{Wenger}}, \bibnamefont{and}
  \bibinfo{author}{\bibfnamefont{D.~S.} \bibnamefont{Weiss}},
  \bibinfo{journal}{Science} \textbf{\bibinfo{volume}{305}},
  \bibinfo{pages}{1125} (\bibinfo{year}{2004}).

\bibitem[{\citenamefont{Kinoshita et~al.}(2005)\citenamefont{Kinoshita, Wenger,
  and Weiss}}]{Kinoshita2005}
\bibinfo{author}{\bibfnamefont{T.}~\bibnamefont{Kinoshita}},
  \bibinfo{author}{\bibfnamefont{T.}~\bibnamefont{Wenger}}, \bibnamefont{and}
  \bibinfo{author}{\bibfnamefont{D.~S.} \bibnamefont{Weiss}},
  \bibinfo{journal}{Phys. Rev. Lett.} \textbf{\bibinfo{volume}{95}},
  \bibinfo{eid}{190406}  (\bibinfo{year}{2005}).

\bibitem[{\citenamefont{Kinoshita et~al.}(2006)\citenamefont{Kinoshita, Wenger,
  and Weiss}}]{Kinoshita2006}
\bibinfo{author}{\bibfnamefont{T.}~\bibnamefont{Kinoshita}},
  \bibinfo{author}{\bibfnamefont{T.}~\bibnamefont{Wenger}}, \bibnamefont{and}
  \bibinfo{author}{\bibfnamefont{D.~S.} \bibnamefont{Weiss}},
  \bibinfo{journal}{Nature} \textbf{\bibinfo{volume}{440}},
  \bibinfo{pages}{900} (\bibinfo{year}{2006}).

\bibitem[{\citenamefont{Tolra et~al.}(2004)\citenamefont{Tolra, O'Hara,
  Huckans, Phillips, Rolston, and Porto}}]{Tolra2004}
\bibinfo{author}{\bibfnamefont{B.~L.} \bibnamefont{Tolra}},
  \bibinfo{author}{\bibfnamefont{K.~M.} \bibnamefont{O'Hara}},
  \bibinfo{author}{\bibfnamefont{J.~H.} \bibnamefont{Huckans}},
  \bibinfo{author}{\bibfnamefont{W.~D.} \bibnamefont{Phillips}},
  \bibinfo{author}{\bibfnamefont{S.~L.} \bibnamefont{Rolston}},
  \bibnamefont{and} \bibinfo{author}{\bibfnamefont{J.~V.} \bibnamefont{Porto}},
  \bibinfo{journal}{Phys. Rev. Lett.} \textbf{\bibinfo{volume}{92}},
  \bibinfo{eid}{190401}  (\bibinfo{year}{2004}).

\bibitem[{\citenamefont{Fertig et~al.}(2005)\citenamefont{Fertig, O'Hara,
  Huckans, Rolston, Phillips, and Porto}}]{Fertig2005}
\bibinfo{author}{\bibfnamefont{C.~D.} \bibnamefont{Fertig}},
  \bibinfo{author}{\bibfnamefont{K.~M.} \bibnamefont{O'Hara}},
  \bibinfo{author}{\bibfnamefont{J.~H.} \bibnamefont{Huckans}},
  \bibinfo{author}{\bibfnamefont{S.~L.} \bibnamefont{Rolston}},
  \bibinfo{author}{\bibfnamefont{W.~D.} \bibnamefont{Phillips}},
  \bibnamefont{and} \bibinfo{author}{\bibfnamefont{J.~V.} \bibnamefont{Porto}},
  \bibinfo{journal}{Phys. Rev. Lett.} \textbf{\bibinfo{volume}{94}},
  \bibinfo{eid}{120403}  (\bibinfo{year}{2005}).

\bibitem[{\citenamefont{Yukalov}(2009)}]{Yukalov2009}
\bibinfo{author}{\bibfnamefont{V.~I.} \bibnamefont{Yukalov}},
  \bibinfo{journal}{Laser Phys.} \textbf{\bibinfo{volume}{19}},
  \bibinfo{pages}{1} (\bibinfo{year}{2009}).

\bibitem[{\citenamefont{G\"orlitz et~al.}(2001)\citenamefont{G\"orlitz, Vogels,
  Leanhardt, Raman, Gustavson, Abo-Shaeer, Chikkatur, Gupta, Inouye, Rosenband
  et~al.}}]{Gorlitz2001}
\bibinfo{author}{\bibfnamefont{A.}~\bibnamefont{G\"orlitz}},
  \bibinfo{author}{\bibfnamefont{J.~M.} \bibnamefont{Vogels}},
  \bibinfo{author}{\bibfnamefont{A.~E.} \bibnamefont{Leanhardt}},
  \bibinfo{author}{\bibfnamefont{C.}~\bibnamefont{Raman}},
  \bibinfo{author}{\bibfnamefont{T.~L.} \bibnamefont{Gustavson}},
  \bibinfo{author}{\bibfnamefont{J.~R.} \bibnamefont{Abo-Shaeer}},
  \bibinfo{author}{\bibfnamefont{A.~P.} \bibnamefont{Chikkatur}},
  \bibinfo{author}{\bibfnamefont{S.}~\bibnamefont{Gupta}},
  \bibinfo{author}{\bibfnamefont{S.}~\bibnamefont{Inouye}},
  \bibinfo{author}{\bibfnamefont{T.}~\bibnamefont{Rosenband}},
  \bibnamefont{et~al.}, \bibinfo{journal}{Phys. Rev. Lett.}
  \textbf{\bibinfo{volume}{87}}, \bibinfo{pages}{130402}
  (\bibinfo{year}{2001}).

\bibitem[{\citenamefont{Leanhardt et~al.}(2002)\citenamefont{Leanhardt,
  Chikkatur, Kielpinski, Shin, Gustavson, Ketterle, and
  Pritchard}}]{Leanhardt2002}
\bibinfo{author}{\bibfnamefont{A.~E.} \bibnamefont{Leanhardt}},
  \bibinfo{author}{\bibfnamefont{A.~P.} \bibnamefont{Chikkatur}},
  \bibinfo{author}{\bibfnamefont{D.}~\bibnamefont{Kielpinski}},
  \bibinfo{author}{\bibfnamefont{Y.}~\bibnamefont{Shin}},
  \bibinfo{author}{\bibfnamefont{T.~L.} \bibnamefont{Gustavson}},
  \bibinfo{author}{\bibfnamefont{W.}~\bibnamefont{Ketterle}}, \bibnamefont{and}
  \bibinfo{author}{\bibfnamefont{D.~E.} \bibnamefont{Pritchard}},
  \bibinfo{journal}{Phys. Rev. Lett.} \textbf{\bibinfo{volume}{89}},
  \bibinfo{pages}{040401} (\bibinfo{year}{2002}).

\bibitem[{\citenamefont{Strecker et~al.}(2002)\citenamefont{Strecker,
  Partridge, Truscott, and Hulet}}]{Strecker2002}
\bibinfo{author}{\bibfnamefont{K.~E.} \bibnamefont{Strecker}},
  \bibinfo{author}{\bibfnamefont{G.~B.} \bibnamefont{Partridge}},
  \bibinfo{author}{\bibfnamefont{A.~G.} \bibnamefont{Truscott}},
  \bibnamefont{and} \bibinfo{author}{\bibfnamefont{R.~G.} \bibnamefont{Hulet}},
  \bibinfo{journal}{Nature} \textbf{\bibinfo{volume}{417}},
  \bibinfo{pages}{150} (\bibinfo{year}{2002}).

\bibitem[{\citenamefont{Khaykovich et~al.}(2002)\citenamefont{Khaykovich,
  Schreck, Ferrari, Bourdel, Cubizolles, Carr, Castin, and
  Salomon}}]{Khaykovich2002}
\bibinfo{author}{\bibfnamefont{L.}~\bibnamefont{Khaykovich}},
  \bibinfo{author}{\bibfnamefont{F.}~\bibnamefont{Schreck}},
  \bibinfo{author}{\bibfnamefont{G.}~\bibnamefont{Ferrari}},
  \bibinfo{author}{\bibfnamefont{T.}~\bibnamefont{Bourdel}},
  \bibinfo{author}{\bibfnamefont{J.}~\bibnamefont{Cubizolles}},
  \bibinfo{author}{\bibfnamefont{L.~D.} \bibnamefont{Carr}},
  \bibinfo{author}{\bibfnamefont{Y.}~\bibnamefont{Castin}}, \bibnamefont{and}
  \bibinfo{author}{\bibfnamefont{C.}~\bibnamefont{Salomon}},
  \bibinfo{journal}{Science} \textbf{\bibinfo{volume}{296}},
  \bibinfo{pages}{1290} (\bibinfo{year}{2002}).

\bibitem[{\citenamefont{Richard et~al.}(2003)\citenamefont{Richard, Gerbier,
  Thywissen, Hugbart, Bouyer, and Aspect}}]{Richard2003}
\bibinfo{author}{\bibfnamefont{S.}~\bibnamefont{Richard}},
  \bibinfo{author}{\bibfnamefont{F.}~\bibnamefont{Gerbier}},
  \bibinfo{author}{\bibfnamefont{J.~H.} \bibnamefont{Thywissen}},
  \bibinfo{author}{\bibfnamefont{M.}~\bibnamefont{Hugbart}},
  \bibinfo{author}{\bibfnamefont{P.}~\bibnamefont{Bouyer}}, \bibnamefont{and}
  \bibinfo{author}{\bibfnamefont{A.}~\bibnamefont{Aspect}},
  \bibinfo{journal}{Phys. Rev. Lett.} \textbf{\bibinfo{volume}{91}},
  \bibinfo{eid}{010405}  (\bibinfo{year}{2003}).

\bibitem[{\citenamefont{Hugbart et~al.}(2007)\citenamefont{Hugbart, Retter,
  Varon, Bouyer, Aspect, and Davis}}]{Hugbart2007}
\bibinfo{author}{\bibfnamefont{M.}~\bibnamefont{Hugbart}},
  \bibinfo{author}{\bibfnamefont{J.~A.} \bibnamefont{Retter}},
  \bibinfo{author}{\bibfnamefont{A.~F.} \bibnamefont{Varon}},
  \bibinfo{author}{\bibfnamefont{P.}~\bibnamefont{Bouyer}},
  \bibinfo{author}{\bibfnamefont{A.}~\bibnamefont{Aspect}}, \bibnamefont{and}
  \bibinfo{author}{\bibfnamefont{M.~J.} \bibnamefont{Davis}},
  \bibinfo{journal}{Phys. Rev. A} \textbf{\bibinfo{volume}{75}},
  \bibinfo{eid}{011602}  (\bibinfo{year}{2007}).

\bibitem[{\citenamefont{Folman et~al.}(2002)\citenamefont{Folman, Krueger,
  Schmiedmayer, Denschlag, and Henkel}}]{Folman2002}
\bibinfo{author}{\bibfnamefont{R.}~\bibnamefont{Folman}},
  \bibinfo{author}{\bibfnamefont{P.}~\bibnamefont{Krueger}},
  \bibinfo{author}{\bibfnamefont{J.}~\bibnamefont{Schmiedmayer}},
  \bibinfo{author}{\bibfnamefont{J.}~\bibnamefont{Denschlag}},
  \bibnamefont{and} \bibinfo{author}{\bibfnamefont{C.}~\bibnamefont{Henkel}},
  in \emph{\bibinfo{booktitle}{Adv. At. Mol. Opt. Phys.}}
  (\bibinfo{publisher}{Academic Press}, \bibinfo{address}{New York},
  \bibinfo{year}{2002}), vol.~\bibinfo{volume}{48}, pp.
  \bibinfo{pages}{263--356}.

\bibitem[{\citenamefont{Esteve et~al.}(2006)\citenamefont{Esteve, Trebbia,
  Schumm, Aspect, Westbrook, and Bouchoule}}]{Esteve2006}
\bibinfo{author}{\bibfnamefont{J.}~\bibnamefont{Esteve}},
  \bibinfo{author}{\bibfnamefont{J.-B.} \bibnamefont{Trebbia}},
  \bibinfo{author}{\bibfnamefont{T.}~\bibnamefont{Schumm}},
  \bibinfo{author}{\bibfnamefont{A.}~\bibnamefont{Aspect}},
  \bibinfo{author}{\bibfnamefont{C.~I.} \bibnamefont{Westbrook}},
  \bibnamefont{and}
  \bibinfo{author}{\bibfnamefont{I.}~\bibnamefont{Bouchoule}},
  \bibinfo{journal}{Phys. Rev. Lett.} \textbf{\bibinfo{volume}{96}},
  \bibinfo{eid}{130403}  (\bibinfo{year}{2006}).

\bibitem[{\citenamefont{Hofferberth et~al.}(2007)\citenamefont{Hofferberth,
  Lesanovsky, Fischer, Schumm, and Schmiedmayer}}]{Hofferberth2007}
\bibinfo{author}{\bibfnamefont{S.}~\bibnamefont{Hofferberth}},
  \bibinfo{author}{\bibfnamefont{I.}~\bibnamefont{Lesanovsky}},
  \bibinfo{author}{\bibfnamefont{B.}~\bibnamefont{Fischer}},
  \bibinfo{author}{\bibfnamefont{T.}~\bibnamefont{Schumm}}, \bibnamefont{and}
  \bibinfo{author}{\bibfnamefont{J.}~\bibnamefont{Schmiedmayer}},
  \bibinfo{journal}{Nature} \textbf{\bibinfo{volume}{449}},
  \bibinfo{pages}{324} (\bibinfo{year}{2007}).

\bibitem[{\citenamefont{Jo et~al.}(2007)\citenamefont{Jo, Choi, Christensen,
  Lee, Pasquini, Ketterle, and Pritchard}}]{Jo2007}
\bibinfo{author}{\bibfnamefont{G.-B.} \bibnamefont{Jo}},
  \bibinfo{author}{\bibfnamefont{J.-H.} \bibnamefont{Choi}},
  \bibinfo{author}{\bibfnamefont{C.~A.} \bibnamefont{Christensen}},
  \bibinfo{author}{\bibfnamefont{Y.-R.} \bibnamefont{Lee}},
  \bibinfo{author}{\bibfnamefont{T.~A.} \bibnamefont{Pasquini}},
  \bibinfo{author}{\bibfnamefont{W.}~\bibnamefont{Ketterle}}, \bibnamefont{and}
  \bibinfo{author}{\bibfnamefont{D.~E.} \bibnamefont{Pritchard}},
  \bibinfo{journal}{Phys. Rev. Lett.} \textbf{\bibinfo{volume}{99}},
  \bibinfo{eid}{240406}  (\bibinfo{year}{2007}).

\bibitem[{\citenamefont{van Amerongen et~al.}(2008)\citenamefont{van Amerongen,
  van Es, Wicke, Kheruntsyan, and van Druten}}]{Amerongen2008}
\bibinfo{author}{\bibfnamefont{A.~H.} \bibnamefont{van Amerongen}},
  \bibinfo{author}{\bibfnamefont{J.~J.~P.} \bibnamefont{van Es}},
  \bibinfo{author}{\bibfnamefont{P.}~\bibnamefont{Wicke}},
  \bibinfo{author}{\bibfnamefont{K.~V.} \bibnamefont{Kheruntsyan}},
  \bibnamefont{and} \bibinfo{author}{\bibfnamefont{N.~J.} \bibnamefont{van
  Druten}}, \bibinfo{journal}{Phys. Rev. Lett.} \textbf{\bibinfo{volume}{100}},
  \bibinfo{eid}{090402}  (\bibinfo{year}{2008}).

\bibitem[{\citenamefont{Yurovsky et~al.}({2008})\citenamefont{Yurovsky,
  Olshanii, and Weiss}}]{Yurovsky2008b}
\bibinfo{author}{\bibfnamefont{V.~A.} \bibnamefont{Yurovsky}},
  \bibinfo{author}{\bibfnamefont{M.}~\bibnamefont{Olshanii}}, \bibnamefont{and}
  \bibinfo{author}{\bibfnamefont{D.~S.} \bibnamefont{Weiss}}, in
  \emph{\bibinfo{booktitle}{Adv. At. Mol. Opt. Phys.}}
  (\bibinfo{publisher}{Elsvier Academic Press}, \bibinfo{address}{New York},
  \bibinfo{year}{{2008}}), vol.~\bibinfo{volume}{{55}}, pp.
  \bibinfo{pages}{{61--138}}.

\bibitem[{\citenamefont{Lieb and Liniger}(1963)}]{Lieb1963}
\bibinfo{author}{\bibfnamefont{E.~H.} \bibnamefont{Lieb}} \bibnamefont{and}
  \bibinfo{author}{\bibfnamefont{W.}~\bibnamefont{Liniger}},
  \bibinfo{journal}{Phys. Rev.} \textbf{\bibinfo{volume}{130}},
  \bibinfo{pages}{1605} (\bibinfo{year}{1963}).

\bibitem[{\citenamefont{McGuire}(1964)}]{McGuire1964}
\bibinfo{author}{\bibfnamefont{J.~B.} \bibnamefont{McGuire}},
  \bibinfo{journal}{J. Math. Phys.} \textbf{\bibinfo{volume}{5}},
  \bibinfo{pages}{622} (\bibinfo{year}{1964}).

\bibitem[{\citenamefont{Berezin et~al.}(1964)\citenamefont{Berezin, Pohil, and
  Finkelberg}}]{Berezin1964}
\bibinfo{author}{\bibfnamefont{F.~A.} \bibnamefont{Berezin}},
  \bibinfo{author}{\bibfnamefont{G.~P.} \bibnamefont{Pohil}}, \bibnamefont{and}
  \bibinfo{author}{\bibfnamefont{V.~M.} \bibnamefont{Finkelberg}},
  \bibinfo{journal}{Vestnik Moskovskogo Universiteta (in Russian)}
  \textbf{\bibinfo{volume}{No. 1}}, \bibinfo{pages}{21} (\bibinfo{year}{1964}).

\bibitem[{\citenamefont{Dodd}(1972)}]{Dodd1972}
\bibinfo{author}{\bibfnamefont{L.~R.} \bibnamefont{Dodd}},
  \bibinfo{journal}{Australian Journal of Physics}
  \textbf{\bibinfo{volume}{25}}, \bibinfo{pages}{507} (\bibinfo{year}{1972}).

\bibitem[{\citenamefont{Bergeman et~al.}(2003)\citenamefont{Bergeman, Moore,
  and Olshanii}}]{Bergeman2003}
\bibinfo{author}{\bibfnamefont{T.}~\bibnamefont{Bergeman}},
  \bibinfo{author}{\bibfnamefont{M.~G.} \bibnamefont{Moore}}, \bibnamefont{and}
  \bibinfo{author}{\bibfnamefont{M.}~\bibnamefont{Olshanii}},
  \bibinfo{journal}{Phys. Rev. Lett.} \textbf{\bibinfo{volume}{91}},
  \bibinfo{eid}{163201}  (\bibinfo{year}{2003}).

\bibitem[{\citenamefont{Moore et~al.}(2004)\citenamefont{Moore, Bergeman, and
  Olshanii}}]{Moore2004}
\bibinfo{author}{\bibfnamefont{M.~G.} \bibnamefont{Moore}},
  \bibinfo{author}{\bibfnamefont{T.}~\bibnamefont{Bergeman}}, \bibnamefont{and}
  \bibinfo{author}{\bibfnamefont{M.}~\bibnamefont{Olshanii}},
  \bibinfo{journal}{J. Phys. IV (France)} \textbf{\bibinfo{volume}{116}},
  \bibinfo{pages}{69} (\bibinfo{year}{2004}).

\bibitem[{\citenamefont{Timmermans et~al.}(1999)\citenamefont{Timmermans,
  Tommasini, Hussein, and Kerman}}]{Timmermans1999}
\bibinfo{author}{\bibfnamefont{E.}~\bibnamefont{Timmermans}},
  \bibinfo{author}{\bibfnamefont{P.}~\bibnamefont{Tommasini}},
  \bibinfo{author}{\bibfnamefont{M.}~\bibnamefont{Hussein}}, \bibnamefont{and}
  \bibinfo{author}{\bibfnamefont{A.}~\bibnamefont{Kerman}},
  \bibinfo{journal}{Phys. Rep.} \textbf{\bibinfo{volume}{315}},
  \bibinfo{pages}{199} (\bibinfo{year}{1999}).

\bibitem[{\citenamefont{Yurovsky}(2005)}]{Yurovsky2005}
\bibinfo{author}{\bibfnamefont{V.~A.} \bibnamefont{Yurovsky}},
  \bibinfo{journal}{Phys. Rev. A} \textbf{\bibinfo{volume}{71}},
  \bibinfo{eid}{012709}  (\bibinfo{year}{2005}).

\bibitem[{\citenamefont{Yurovsky}(2006)}]{Yurovsky2006b}
\bibinfo{author}{\bibfnamefont{V.~A.} \bibnamefont{Yurovsky}},
  \bibinfo{journal}{Phys. Rev. A} \textbf{\bibinfo{volume}{73}},
  \bibinfo{eid}{052709}  (\bibinfo{year}{2006})
  .

\bibitem[{\citenamefont{Kim et~al.}(2005)\citenamefont{Kim, Schmiedmayer, and
  Schmelcher}}]{Kim2005}
\bibinfo{author}{\bibfnamefont{J.~I.} \bibnamefont{Kim}},
  \bibinfo{author}{\bibfnamefont{J.}~\bibnamefont{Schmiedmayer}},
  \bibnamefont{and}
  \bibinfo{author}{\bibfnamefont{P.}~\bibnamefont{Schmelcher}},
  \bibinfo{journal}{Phys. Rev. A} \textbf{\bibinfo{volume}{72}},
  \bibinfo{eid}{042711}  (\bibinfo{year}{2005}).

\bibitem[{\citenamefont{Peano et~al.}(2005{\natexlab{a}})\citenamefont{Peano,
  Thorwart, Mora, and Egger}}]{Peano2005}
\bibinfo{author}{\bibfnamefont{V.}~\bibnamefont{Peano}},
  \bibinfo{author}{\bibfnamefont{M.}~\bibnamefont{Thorwart}},
  \bibinfo{author}{\bibfnamefont{C.}~\bibnamefont{Mora}}, \bibnamefont{and}
  \bibinfo{author}{\bibfnamefont{R.}~\bibnamefont{Egger}},
  \bibinfo{journal}{New J. Phys.} \textbf{\bibinfo{volume}{7}},
  \bibinfo{pages}{192} (\bibinfo{year}{2005}{\natexlab{a}}).

\bibitem[{\citenamefont{Peano et~al.}(2005{\natexlab{b}})\citenamefont{Peano,
  Thorwart, Kasper, and Egger}}]{Peano2005a}
\bibinfo{author}{\bibfnamefont{V.}~\bibnamefont{Peano}},
  \bibinfo{author}{\bibfnamefont{M.}~\bibnamefont{Thorwart}},
  \bibinfo{author}{\bibfnamefont{A.}~\bibnamefont{Kasper}}, \bibnamefont{and}
  \bibinfo{author}{\bibfnamefont{R.}~\bibnamefont{Egger}},
  \bibinfo{journal}{Appl. Phys. B} \textbf{\bibinfo{volume}{81}},
  \bibinfo{pages}{1075} (\bibinfo{year}{2005}{\natexlab{b}}).

\bibitem[{\citenamefont{Kim et~al.}(2006)\citenamefont{Kim, Melezhik, and
  Schmelcher}}]{Kim2006}
\bibinfo{author}{\bibfnamefont{J.~I.} \bibnamefont{Kim}},
  \bibinfo{author}{\bibfnamefont{V.~S.} \bibnamefont{Melezhik}},
  \bibnamefont{and}
  \bibinfo{author}{\bibfnamefont{P.}~\bibnamefont{Schmelcher}},
  \bibinfo{journal}{Phys. Rev. Lett.} \textbf{\bibinfo{volume}{97}},
  \bibinfo{eid}{193203}  (\bibinfo{year}{2006}).

\bibitem[{\citenamefont{Naidon et~al.}(2007)\citenamefont{Naidon, Tiesinga,
  Mitchell, and Julienne}}]{Naidon2007}
\bibinfo{author}{\bibfnamefont{P.}~\bibnamefont{Naidon}},
  \bibinfo{author}{\bibfnamefont{E.}~\bibnamefont{Tiesinga}},
  \bibinfo{author}{\bibfnamefont{W.~F.} \bibnamefont{Mitchell}},
  \bibnamefont{and} \bibinfo{author}{\bibfnamefont{P.~S.}
  \bibnamefont{Julienne}}, \bibinfo{journal}{New J. Phys.}
  \textbf{\bibinfo{volume}{9}}, \bibinfo{pages}{19} (\bibinfo{year}{2007}).

\bibitem[{\citenamefont{Melezhik et~al.}(2007)\citenamefont{Melezhik, Kim, and
  Schmelcher}}]{Melezhik2007}
\bibinfo{author}{\bibfnamefont{V.~S.} \bibnamefont{Melezhik}},
  \bibinfo{author}{\bibfnamefont{J.~I.} \bibnamefont{Kim}}, \bibnamefont{and}
  \bibinfo{author}{\bibfnamefont{P.}~\bibnamefont{Schmelcher}},
  \bibinfo{journal}{Phys. Rev. A} \textbf{\bibinfo{volume}{76}},
  \bibinfo{eid}{053611}  (\bibinfo{year}{2007}).

\bibitem[{\citenamefont{Saeidian et~al.}(2008)\citenamefont{Saeidian, Melezhik,
  and Schmelcher}}]{Saeidian2008}
\bibinfo{author}{\bibfnamefont{S.}~\bibnamefont{Saeidian}},
  \bibinfo{author}{\bibfnamefont{V.~S.} \bibnamefont{Melezhik}},
  \bibnamefont{and}
  \bibinfo{author}{\bibfnamefont{P.}~\bibnamefont{Schmelcher}},
  \bibinfo{journal}{Phys. Rev. A} \textbf{\bibinfo{volume}{77}},
  \bibinfo{eid}{042721}  (\bibinfo{year}{2008}).

\bibitem[{\citenamefont{Mora et~al.}(2005)\citenamefont{Mora, Egger, and
  Gogolin}}]{Mora2005}
\bibinfo{author}{\bibfnamefont{C.}~\bibnamefont{Mora}},
  \bibinfo{author}{\bibfnamefont{R.}~\bibnamefont{Egger}}, \bibnamefont{and}
  \bibinfo{author}{\bibfnamefont{A.~O.} \bibnamefont{Gogolin}},
  \bibinfo{journal}{Phys. Rev. A} \textbf{\bibinfo{volume}{71}},
  \bibinfo{eid}{052705}  (\bibinfo{year}{2005}).

\bibitem[{\citenamefont{Sinha et~al.}(2006)\citenamefont{Sinha, Cherny,
  Kovrizhin, and Brand}}]{Sinha2006}
\bibinfo{author}{\bibfnamefont{S.}~\bibnamefont{Sinha}},
  \bibinfo{author}{\bibfnamefont{A.~Y.} \bibnamefont{Cherny}},
  \bibinfo{author}{\bibfnamefont{D.}~\bibnamefont{Kovrizhin}},
  \bibnamefont{and} \bibinfo{author}{\bibfnamefont{J.}~\bibnamefont{Brand}},
  \bibinfo{journal}{Phys. Rev. Lett.} \textbf{\bibinfo{volume}{96}},
  \bibinfo{eid}{030406}  (\bibinfo{year}{2006}).

\bibitem[{\citenamefont{Mazets et~al.}(2008)\citenamefont{Mazets, Schumm, and
  Schmiedmayer}}]{Mazets2008}
\bibinfo{author}{\bibfnamefont{I.~E.} \bibnamefont{Mazets}},
  \bibinfo{author}{\bibfnamefont{T.}~\bibnamefont{Schumm}}, \bibnamefont{and}
  \bibinfo{author}{\bibfnamefont{J.}~\bibnamefont{Schmiedmayer}},
  \bibinfo{journal}{Phys. Rev. Lett.} \textbf{\bibinfo{volume}{100}},
  \bibinfo{eid}{210403}  (\bibinfo{year}{2008}).

\bibitem[{\citenamefont{Yurovsky et~al.}(2006)\citenamefont{Yurovsky,
  Ben-Reuven, and Olshanii}}]{Yurovsky2006}
\bibinfo{author}{\bibfnamefont{V.~A.} \bibnamefont{Yurovsky}},
  \bibinfo{author}{\bibfnamefont{A.}~\bibnamefont{Ben-Reuven}},
  \bibnamefont{and} \bibinfo{author}{\bibfnamefont{M.}~\bibnamefont{Olshanii}},
  \bibinfo{journal}{Phys. Rev. Lett.} \textbf{\bibinfo{volume}{96}},
  \bibinfo{eid}{163201}  (\bibinfo{year}{2006}).

\bibitem[{\citenamefont{Yurovsky}(2008)}]{Yurovsky2008}
\bibinfo{author}{\bibfnamefont{V.~A.} \bibnamefont{Yurovsky}},
  \bibinfo{journal}{Phys. Rev. A} \textbf{\bibinfo{volume}{77}},
  \bibinfo{eid}{012716}  (\bibinfo{year}{2008}).

\bibitem[{\citenamefont{Schmid and Ziegelmann}(1974)}]{Schmid}
\bibinfo{author}{\bibfnamefont{E.~W.} \bibnamefont{Schmid}} \bibnamefont{and}
  \bibinfo{author}{\bibfnamefont{H.}~\bibnamefont{Ziegelmann}},
  \emph{\bibinfo{title}{{The Quantum Mechanical Three-Body Problem}}}
  (\bibinfo{publisher}{Pergamon Press}, \bibinfo{address}{Oxford},
  \bibinfo{year}{1974}).

\bibitem[{\citenamefont{Gl{\"o}ckle}(1983)}]{Glockle}
\bibinfo{author}{\bibfnamefont{W.}~\bibnamefont{Gl{\"o}ckle}},
  \emph{\bibinfo{title}{Quantim Mechanical Few-Body Problem}}
  (\bibinfo{publisher}{Springer}, \bibinfo{address}{Berlin},
  \bibinfo{year}{1983}).

\bibitem[{\citenamefont{Dodd}(1970)}]{Dodd1970}
\bibinfo{author}{\bibfnamefont{L.~R.} \bibnamefont{Dodd}}, \bibinfo{journal}{J.
  Math. Phys.} \textbf{\bibinfo{volume}{11}}, \bibinfo{pages}{207}
  (\bibinfo{year}{1970}).

\bibitem[{\citenamefont{Majumdar}(1972)}]{Majumdar1972}
\bibinfo{author}{\bibfnamefont{C.~K.} \bibnamefont{Majumdar}},
  \bibinfo{journal}{J. Math. Phys.} \textbf{\bibinfo{volume}{13}},
  \bibinfo{pages}{705} (\bibinfo{year}{1972}).

\bibitem[{\citenamefont{Mehta and Shepard}(2005)}]{Mehta2005}
\bibinfo{author}{\bibfnamefont{N.~P.} \bibnamefont{Mehta}} \bibnamefont{and}
  \bibinfo{author}{\bibfnamefont{J.~R.} \bibnamefont{Shepard}},
  \bibinfo{journal}{Phys. Rev. A} \textbf{\bibinfo{volume}{72}},
  \bibinfo{eid}{032728}  (\bibinfo{year}{2005}).

\bibitem[{\citenamefont{Melde et~al.}(2002)\citenamefont{Melde, Canton, and
  Svenne}}]{Melde2002}
\bibinfo{author}{\bibfnamefont{T.}~\bibnamefont{Melde}},
  \bibinfo{author}{\bibfnamefont{L.}~\bibnamefont{Canton}}, \bibnamefont{and}
  \bibinfo{author}{\bibfnamefont{J.}~\bibnamefont{Svenne}},
  \bibinfo{journal}{Few-Body Systems} \textbf{\bibinfo{volume}{32}},
  \bibinfo{pages}{143} (\bibinfo{year}{2002}).

\bibitem[{\citenamefont{Mehta et~al.}(2007)\citenamefont{Mehta, Esry, and
  Greene}}]{Mehta2007}
\bibinfo{author}{\bibfnamefont{N.~P.} \bibnamefont{Mehta}},
  \bibinfo{author}{\bibfnamefont{B.~D.} \bibnamefont{Esry}}, \bibnamefont{and}
  \bibinfo{author}{\bibfnamefont{C.~H.} \bibnamefont{Greene}},
  \bibinfo{journal}{Phys. Rev. A} \textbf{\bibinfo{volume}{76}},
  \bibinfo{eid}{022711}  (\bibinfo{year}{2007}).

\bibitem[{\citenamefont{Kheruntsyan and Drummond}(1998)}]{Kheruntsyan1998}
\bibinfo{author}{\bibfnamefont{K.~V.} \bibnamefont{Kheruntsyan}}
  \bibnamefont{and} \bibinfo{author}{\bibfnamefont{P.~D.}
  \bibnamefont{Drummond}}, \bibinfo{journal}{Phys. Rev. A}
  \textbf{\bibinfo{volume}{58}}, \bibinfo{pages}{2488} (\bibinfo{year}{1998}).

\bibitem[{\citenamefont{Abdurakhmanov and Zubarev}(1985)}]{Abdurakhmanov1985}
\bibinfo{author}{\bibfnamefont{A.}~\bibnamefont{Abdurakhmanov}}
  \bibnamefont{and} \bibinfo{author}{\bibfnamefont{A.~L.}
  \bibnamefont{Zubarev}}, \bibinfo{journal}{Z. Phys. A}
  \textbf{\bibinfo{volume}{322}}, \bibinfo{pages}{523} (\bibinfo{year}{1985}).

\bibitem[{\citenamefont{Abdurakhmanov et~al.}(1987)\citenamefont{Abdurakhmanov,
  Zubarev, Latipov, and Nasyrov}}]{Abdurakhmanov1987}
\bibinfo{author}{\bibfnamefont{A.}~\bibnamefont{Abdurakhmanov}},
  \bibinfo{author}{\bibfnamefont{A.~L.} \bibnamefont{Zubarev}},
  \bibinfo{author}{\bibfnamefont{A.~S.} \bibnamefont{Latipov}},
  \bibnamefont{and} \bibinfo{author}{\bibfnamefont{M.}~\bibnamefont{Nasyrov}},
  \bibinfo{journal}{Sov. J. Nucl. Phys.} \textbf{\bibinfo{volume}{46}},
  \bibinfo{pages}{217} (\bibinfo{year}{1987}).

\bibitem[{\citenamefont{Vinitskii et~al.}(1992)\citenamefont{Vinitskii,
  Kuperin, Motovilov, and Suz'ko}}]{Vinitskii1992}
\bibinfo{author}{\bibfnamefont{S.}~\bibnamefont{Vinitskii}},
  \bibinfo{author}{\bibfnamefont{Y.}~\bibnamefont{Kuperin}},
  \bibinfo{author}{\bibfnamefont{A.}~\bibnamefont{Motovilov}},
  \bibnamefont{and} \bibinfo{author}{\bibfnamefont{A.}~\bibnamefont{Suz'ko}},
  \bibinfo{journal}{Sov. J. Nucl. Phys.} \textbf{\bibinfo{volume}{55}},
  \bibinfo{pages}{245} (\bibinfo{year}{1992}).

\bibitem[{\citenamefont{Lovelace}(1964)}]{Lovelace1964}
\bibinfo{author}{\bibfnamefont{C.}~\bibnamefont{Lovelace}},
  \bibinfo{journal}{Phys. Rev.} \textbf{\bibinfo{volume}{135}},
  \bibinfo{pages}{B1225} (\bibinfo{year}{1964}).

\bibitem[{\citenamefont{Fedichev et~al.}(1996)\citenamefont{Fedichev, Reynolds,
  and Shlyapnikov}}]{Fedichev1996b}
\bibinfo{author}{\bibfnamefont{P.~O.} \bibnamefont{Fedichev}},
  \bibinfo{author}{\bibfnamefont{M.~W.} \bibnamefont{Reynolds}},
  \bibnamefont{and} \bibinfo{author}{\bibfnamefont{G.~V.}
  \bibnamefont{Shlyapnikov}}, \bibinfo{journal}{Phys. Rev. Lett.}
  \textbf{\bibinfo{volume}{77}}, \bibinfo{pages}{2921} (\bibinfo{year}{1996}).

\bibitem[{\citenamefont{Lipkin}(1973)}]{Lipkin}
\bibinfo{author}{\bibfnamefont{H.~J.} \bibnamefont{Lipkin}},
  \emph{\bibinfo{title}{Quantum Mechanics: New Approaches to Selected Topics}}
  (\bibinfo{publisher}{North-Holland}, \bibinfo{address}{Amsterdam},
  \bibinfo{year}{1973}).

\end{thebibliography}
\end{document}